\documentclass{article}
\RequirePackage{natbib}
\usepackage{graphicx,epsfig,emulateapj,apjfonts}
\newcommand{\Ni}{\ensuremath{^{56}\mathrm{Ni}~}}
\newcommand{\Coa}{\ensuremath{^{56}\mathrm{Co}~}}
\newcommand{\Cob}{\ensuremath{^{57}\mathrm{Co}~}}
\newcommand{\Ti}{\ensuremath{^{44}\mathrm{Ti}~}}
\newcommand{\Kelv}{$^{\circ}\mathrm{K}$}
\renewcommand{\sun}{\odot}
\newcommand{\e}{{\rm e}}
\newcommand{\ergss}{ergs$\;\mbox{s}^{-1}$}
\newcommand{\ergsgm}{ergs$\;\mbox{gm}^{-1}$}
\newcommand{\cms}{cm$\;\mbox{s}^{-1}$}
\newcommand{\msolyr}{$M_\sun\;\mbox{yr}^{-1}$}
\newcommand{\gmcmc}{gm$\;\mbox{cm}^{-3}$}
\newenvironment{inlinetable}{%
\def\@captype{table}%
\noindent\begin{minipage}{0.999\linewidth}\begin{center}\footnotesize}
{\end{center}\end{minipage}\smallskip}

\newenvironment{inlinefigure}{%
\def\@captype{figure}%
\noindent\begin{minipage}{0.999\linewidth}\begin{center}}
{\end{center}\end{minipage}\smallskip}

\begin{document}

\submitted{To appear in the Astrophysical Journal, 540, 2000 September 10}

\title{Black Hole Emergence in Supernovae}

\author{Shmuel Balberg\altaffilmark{1}, Luca Zampieri\altaffilmark{1,2} and 
Stuart L. Shapiro\altaffilmark{1,3}}


\altaffiltext{1}{Department of Physics, Loomis Laboratory of Physics,
University of Illinois at Urbana--Champaign, 1110 West Green Street,
Urbana, IL 61801--3080, sbalberg@astro.physics.uiuc.edu, 
shapiro@astro.physics.uiuc.edu}
\altaffiltext{2}{Department of Physics, University of Padova,
Via Marzolo 8, 35131 Padova, Italy, zampieri@pd.infn.it}
\altaffiltext{3}{Department of Astronomy and National Center for
Supercomputing Applications, University of Illinois
at Urbana--Champaign, Urbana, IL 61801}

\vspace{0.3cm}
                                      
\begin{abstract}

If a black hole formed in a core-collapse supernova is accreting material
from the base of the envelope, the accretion luminosity could be
observable in the supernova light curve. Here we continue the
study of matter fall back onto a black hole
in the wake of a supernova and examine realistic supernovae
models which allow for an early emergence of the accretion luminosity.
Such cases may provide a direct {\it observational} 
identification of the black hole formed in the aftermath of the explosion.
Our approach combines analytic estimates and
fully relativistic, radiation-hydrodynamic numerical computations.
We employ a numerical hydrodynamical scaling technique to accommodate the 
diverse range of dynamical time scales in a single simulation. 
We find that while in typical Type II supernovae heating by radioactive
decays dominates the late-time light curve, low-energy explosions
of more massive stars should provide an important exception where the
accretion luminosity will emerge while it is still relatively large.
Our main focus is on the only current candidate for such an
observation, the very unusual SN1997D. Due to the low energy of the
explosion and the very small ($2\!\times\!10^{-3}\;M_\sun$) inferred
mass of \Coa in the ejected envelope, we find that accretion should become
the dominant source of its luminosity during the year 2000. The
total luminosity at emergence is expected to lie in the range
$0.5-3\!\times\!10^{36}\;$\ergss, potentially detectable with
HST. We also discuss the more favorable case of explosions which eject
negligible amounts of radioactive isotopes and find that
the black hole is likely to emerge a few tens of days
after the explosion, with a luminosity of $\sim 10^{37}\;$\ergss.

\end{abstract}


\vspace{0.3cm}

\section{Introduction}\label{Sect:Intro}

Core-collapse supernovae mark the death of a massive star and the birth 
of a compact remnant. Theory suggests that the remnant can be either a 
neutron star or a black hole, depending on the character 
of the progenitor and the details of the explosion. Radio pulsar 
emission has allowed to compile a variety of observational evidence 
associating neutron stars with sites of known supernovae, but similar 
evidence for a black hole - supernova connection is still mostly 
unavailable. Interestingly, indirect evidence that the black hole 
candidate in the X-ray binary system
GRO J1655-40 was formed in a supernova explosion has recently 
been reported by 
\citet{Israelianal99}. The inference is based on detection of high 
abundances of nitrogen and oxygen on the surface of the companion, which 
has too low a mass to have produced them by thermonuclear burning, hence 
indicating that they were deposited there by the supernova that created 
the black hole. 

The formation of a compact object in a supernova can be inferred 
{\it directly} if the object causes an observable effect on the total 
luminosity which follows the explosion -- i.e., the {\it light curve}. 
In particular, some material from the base of the expanding envelope may 
remain bound to the compact 
object and continuously fall back onto it, thus generating an accretion 
luminosity. \citet{PaperI} have recently performed a 
self-consistent investigation of the accretion luminosity generated by 
spherically symmetric ``fallback'' of matter onto a black hole in the wake of 
a supernova. With the aid of a general-relativistic, radiation 
hydrodynamic Lagrangian code they showed that, at late times, the
fallback evolves quasi-stationarily with the accretion luminosity
obeying the analytic expression
found by \citet{Blondin86} for steady state, spherical hypercritical 
accretion\footnote{We use the term ``hypercritical'' to describe an accretion 
rate $\dot{M}$ that largely exceeds the Eddington limit for the accreting 
object, i.e., $\dot{M}\gg \dot{M}_{Edd}\equiv L_{Edd}/c^2$. 
The actual luminosity that arises would 
still be sub-Eddington if the efficiency of converting energy into 
radiative luminosity is low enough.}. 
The accretion luminosity declines secularly with a time dependence 
of $t^{-25/18}$. This decay is driven by
the decrease in the accretion rate due to the continuous 
expansion of the supernova envelope (Colpi, Shapiro \& Wasserman 1996),

Such an accretion luminosity will produce a distinct 
signature on the total light curve if and when it becomes comparable to 
the output of other power sources, namely the initial internal energy of the 
envelope and decays of radioactive isotopes synthesized in the explosion. 
It is well established that
the efficiency of converting spherical accretion onto a black hole 
into radiative luminosity may be quite low (Shapiro 1973; Blondin 1986;
Wandel, Yahil \& Milgrom 1986; Park 1990; Nobili, Turolla \& Zampieri
1991; Zampieri, Miller \& Turolla 1996), and, in general, 
the accretion luminosity will be undetectable in comparison with the 
luminosity generated by radioactive decays. This is particularly 
important when compared to the case of accretion onto a neutron star, 
which is expected to be far more 
efficient in generating an accretion luminosity. For the last decade this 
distinction has been specifically applied to the case of SN1987A: 
\citet{Chevalier89} and \citet{HouckChev91} found that accretion onto a 
newly formed neutron star in SN1987A would generate an accretion luminosity 
at the Eddington-rate ($\ga 10^{38}\;$\ergss) within a few months after the 
explosion, and the luminosity would persist at this rate for several 
years\footnote{It is noteworthy that magnetic dipole emission of a Crab-like 
pulsar would deposit energy in the envelope at a similar rate 
\citep{Woosleyal89}.}. 
The proximity of SN1987A has allowed observations to follow the 
light curve for over 3000 days (to a present luminosity of 
$\sim10^{36}\;$\ergss), and the light curve has been completely 
consistent with heating by decays of \Coa, \Cob and \Ti. 
The absence of any evidence for accretion luminosity in the case of SN1987A 
has been used to argue that the compact remnant in SN1987A is a 
low mass black hole, implying significant consequences regarding the nuclear 
equation of state \citep{BetheBrown95}. It has also been argued, however, 
that this absence could still be consistent with a weakly 
magnetized neutron star, if the accretion flow has become
dynamically unstable (Fryer, Colgate \& Pinto 1999).

A positive indication of the presence of a black hole would be 
an actual detection of the accretion luminosity. 
The power-law time dependence would make the accretion 
luminosity easily discernible from luminosity generated by radioactive 
heating, which decreases exponentially with time. 
Furthermore, the difference in the time dependence implies that the 
accretion luminosity {\it must} eventually become dominant over  
radioactive decays, so in principle, the black hole always ``emerges'' in 
supernovae light curves. However, the potential for observation is usually 
quite slim: for example, \citet{PaperI} estimated that if a 
low mass black hole was formed in SN1987A, the accretion luminosity 
will not become dominant over that of heating by \Ti decays until $\sim900$ 
years after the explosion, when the luminosity will have dropped to a mere 
$10^{32}\;$\ergss.
The abundance of radioactive elements and the amount of fallback
in SN1987A are rather typical of core collapse supernovae of low to 
intermediate mass progenitors (12--25 $M_\sun$) so that, if a black hole is 
formed, the ``emergence'' of the accretion luminosity occurs only at such late
times and low luminosities that it is of no observational consequence.

An important exception may exist, however, in the explosions of more massive 
stars. In their survey of supernova explosions and nucleosynthesis, 
\citet{WoosWeav95} find that more massive stars tend to produce 
more massive compact remnants: the original iron core is larger, leading to  
a more massive proto-neutron star and to larger amounts of fallback 
(see also Fryer 1999). 
Correspondingly, larger mass progenitors are more likely to produce black 
holes (as the remnant mass would be larger than the maximum mass of stable 
neutron stars), and to have larger amounts of material that remain bound to 
the black hole as a reservoir for late time fallback. Just as important, the 
final abundance of the key radioactive isotopes (\Ni, $^{57}$Ni and \Ti) in 
the expanding envelope is quite sensitive to the location of the cutoff 
that defines the material which settles on the black hole during 
the early stages (while the explosion is still proceeding). 
For the more massive progenitors the cutoff is further out, and 
since the radioactive elements are synthesized in the innermost 
layers above the proto-neutron star, the total mass of these elements in the 
ejected envelopes of higher mass stars can be significantly smaller than in 
``standard'' supernovae. In fact, progenitor of masses in the 
range $30-40\;M_\sun$, are expected to be practically free of radioactive 
elements that can power the bolometric light curve \citep{WoosWeav95}. 
For such stars, 
black hole accretion luminosity should emerge immediately after the decay 
of the recombination peak (tens of days after the explosions), while it is 
still relatively powerful, and hence potentially observable.

Naturally, the more massive stars are a small minority in the stellar 
population, as are their explosions among core collapse supernovae.
However, one recent supernova, SN1997D in NGC 1536 \citep{deMello97}, may 
present a border-line candidate for a direct detection of black hole 
emergence. Its observed light curve indicates that that the total mass of 
\Coa (the daughter of \Ni) in the envelope 
is only $\sim2\!\times\!10^{-3}\;M_\sun$, a factor of $\sim35$ lower than 
SN1987A. By analyzing the light curve and spectra, \citet{Turatto98} 
suggested that SN1997D was a low energy 
($E_{tot}\sim 4\!\times\!10^{50}\;$ergs) 
explosion of a $26\;M_\sun$ star. The expected 
mass of the remnant is about $3\;M_\sun$ 
(formed by a $1.2\;M_\sun$ early fallback on a $1.8\;M_\sun$ collapsed 
core), hence most likely a black hole. Based on the inferred average 
properties of the material available for late time fallback and on the
late time behavior of the accretion luminosity found by \cite{PaperI}, 
Zampieri, Shapiro \& Colpi (1998b) estimated that accretion luminosity will 
emerge in SN1997D in the optical band within {\it only a few years} after 
the explosion.

In this work we revisit the issue of black hole emergence in supernova due to 
spherically-symmetric late-time accretion. In particular, we 
investigate a more robust estimate for our ``proto-type'' case, SN1997D.
We furnish revised analytic 
estimates and a detailed numerical investigation of the 
radiation-hydrodynamic evolution of the ejecta, incorporating the effects
of a realistic chemical composition, opacities and radioactive heating
on the accretion history and luminosity.
The numerical study is based on an improved version 
of the radiation-hydrodynamic code used by \citet{PaperI}, where the main 
modifications are the inclusion of realistic envelope composition and 
opacities (rather than pure hydrogen) and of heating by radioactive decays. 
We find that a $3\;M_\sun$ black hole should emerge in the SN1997D light 
curve at about 1000-1500 days after the explosion -- i.e., {\it during 
late 1999 and 2000} -- when the total bolometric luminosity is 
$\sim 0.5-3\!\times\!10^{36}\;$\ergss, and is marginally detectable with HST.

In the course of our computational study we were compelled to introduce a 
rescaling scheme in order to numerically follow the accretion history and 
light curve for several years. The huge dynamical range of pertinent 
time scales renders a full scale time-integration numerically impractical. 
Our solution has been to study a scaled model which evolves more quickly 
and allows a simple relation between the properties and history of the 
rescaled model and the true one. 
We note that this scheme, which is presented in detail in the appendix, is 
potentially suitable for other problems which involve time-dependent 
radiation hydrodynamics spanning a large dynamic range.
  
We begin our discussion in \S~\ref{Sect:basics} with a summary of the 
characteristic features of supernova light curve evolution when accretion 
onto a central black hole is occurring. 
Quantitative relations between the light curve magnitude and 
the relevant time scales are also reviewed. The principal features of 
the numerical code of \citet{PaperI} and the modifications incorporated for 
this work are described in \S~\ref{Sect:code}. 
The light curve of SN1997D is examined in detail 
in \S~\ref{Sect:SN1997D}, where we present analytic and numerical results, 
including our 
estimates regarding black hole emergence. Prospects of observing black hole 
emergence in other supernovae are discussed in \S~\ref{Sect:otherSN}, and 
further discussion and conclusions are offered in \S~\ref{Sect:conc}.

\section{The Supernova Light Curve including Accretion}\label{Sect:basics}

The bolometric luminosity that follows the explosion of a massive star is
powered by the internal energy of the expanding envelope, emitted as
thermal photons. Under ``bolometric'' we do not include the hard X-ray and
$\gamma-$ray photons that are emitted in radioactive decays and escape from 
the envelope before thermalization. 
In most core collapse supernovae, luminosity
in the early (up to tens of days) light curve is powered by energy
deposited in the envelope by the supernova shock. After this energy is
depleted, only ongoing sources of energy can
continue to power the later part of the light curve, often referred to as
a ``tail''. By definition, radiative 
emission from late fallback qualifies as an ongoing source, as do 
decays of radioactive elements in the envelope\footnote{
\citet{YouBra89} pointed out that if the energy 
generation from radioactive decays is sufficiently high and the progenitor 
radius is sufficiently small, radioactive decays can become the dominant 
source of power for the light curve even very early after the explosion. 
They suggest that this is the case in Type II-linear supernovae.}.

It is useful to examine the evolution of the light curve in
terms of the key characteristic time scales and luminosity scales,
which can be estimated on the basis of the explosion energy 
and the average properties of the progenitor (Arnett 1980, 1996). When
accretion onto the central object caused by late-time fallback is taken
into account, additional time scales and a luminosity scale must also be
included \citep{PaperI}. We briefly review these basic estimates below.
They will serve also as the basis for establishing the rescaling scheme
presented in the appendix.

\subsection{Expansion} \label{subsect:expansion}

For an envelope with initial radius $R_0$ and outer velocity $V_0$ we can
define an initial {\it expansion time scale}, $t_0$, as
\begin{equation}\label{eq:t_exp} 
t_0 = {{R_{0}}\over {V_0}}\;.
\end{equation}
Assuming that after the passage of the shock the envelope settles into
free streaming and homologous expansion (as is typical in strong spherical
shocks), the initial velocity of a shell at radius $r$ in the expanding
envelope is simply $v_0(r)=r/t_0$. For free streaming the radius of the 
shell will follow $r\!\sim\!t$, and since the expansion is nearly adiabatic,  
the temperature $T$ and density $\rho$ of the radiation-pressure dominated 
gas decline according to $T\!\sim\!t^{-1}$ and $\rho\!\sim\!t^{-3}$. These 
quantities determine the leakage rate of thermal photons from the 
envelope material, and hence govern the history of the early light curve.

\subsection{Diffusion} \label{subsect:Diffusion}

During the time that the initial internal energy is the dominant source
for radiative luminosity, the evolution of the light curve is governed by
the properties of the optically-thick, hydrogen-rich outer layer, that
holds most of this energy.  We denote the mass, initial average density
and initial average temperature of this layer as $M_H,\;\rho_{0,H}$ and
$T_{0,H}$, respectively.

The initial post-shock temperature of the envelope is generally high 
enough ($T_{0,H}\ga\!10^6\;$\Kelv) so that the envelope material is
highly ionized, and the dominant source of opacity, $\kappa$, is Thomson
scattering. Correspondingly, the thermal photons are mostly trapped in the
envelope, and a radiative luminosity occurs through diffusion of thermal
energy to the surface. For a homogeneous expanding envelope this diffusion
luminosity can be approximated by \citep{ArnettBook}
 \begin{equation}
 \label{eq:L_diff}
 L_{diff}(t)=L_{diff,0}e^{-(t_0t+t^2/2)/(t_0 t_{diff,0})} \;,
 \end{equation}
where
 \begin{equation}\label{eq:L_diff0}
 L_{diff,0}=\frac{(\frac{4}{3}\pi
 \beta a c)T_{0,H}^4R_0^4}{\kappa M_H}\;,
 \qquad t_{diff,0}=\frac{3\kappa \rho_0 R_0^2}{\beta c}\;,
 \end{equation}
with $\beta$ a numerical factor which depends on the temperature
profile. The value $\beta=13.8$ (exact for the ``radiative-zero''
temperature profile, $T_0(r)\!=\!T_0(\sin(\pi \frac{r}{R_0})/(\pi
\frac{r}{R_0}))^{1/4}$) is usually a reasonable estimate
\citep{Arnett80}. The time dependence of equation~(\ref{eq:L_diff}) includes
the expansion time $t_0$ and the initial diffusion time $t_{diff,0}$,
which is the characteristic time scale for radiation to cross the envelope
through diffusion at the onset of expansion. The characteristic time for
radiation to diffuse out of the {\it expanding} envelope is given by
\citep{PaperI}
 \begin{eqnarray}\label{eq:t_diff}
 t_{diff}\approx  \sqrt{t_0 t_{diff,0}} = 
\left[{1\over {4\pi}} \kappa M_H\right]^{1/2}{1\over {(c V_0)}^{1/2}} \\
 \;\;\; \equiv
 \left[\frac{1}{3}\kappa R_0^3\rho_{0,H}\right]^{1/2}{1\over{(c V_0)}^{1/2}}
  \; .  \nonumber
 \end{eqnarray}

The diffusion approximation breaks down when expansion makes the 
envelope transparent to its thermal photons, after a typical time

\begin{equation}\label{eq:t_trns}
t_{trans} \approx 
\left[{3\over {4\pi}} \kappa M_H\right]^{1/2}{1\over V_0}  \equiv 
[\kappa R_0^3 \rho_{0,H}]^{1/2}{1\over V_0}\;,
\end{equation}
which is significantly longer than $t_{diff}$. However, the onset of 
recombination in the envelope usually occurs early enough so that the actual 
value of $t_{trans}$ is irrelevant.

\subsection{Recombination} \label{subsect:Recombination}

As the temperature of the envelope is gradually degraded by expansion, it 
eventually reaches the characteristic recombination temperature of the 
envelope material. In a hydrogen-rich envelope this temperature 
can be estimated by $T_{rec,H}$ (approximately $10^4\;$\Kelv), so that 
for adiabatic expansion we find a {\it recombination time scale}
 \begin{equation}\label{eq:t_rec}
 t_{rec} = t_0 \frac{T_{0,H}}{T_{rec,H}}\;.
 \end{equation}

The main impact of recombination in the context of the light curve is that 
it imposes a sharp decrease in the opacity to the envelope's own thermal 
photons: as the envelope material recombines, it rapidly transforms from 
opaque to transparent. This process is rapid enough so that the photons 
cannot readjust to the global thermal profile of the envelope; rather, a 
recombination ``front'' sweeps inward through the envelope 
(but outward in the laboratory frame), essentially liberating all the 
thermal energy which is still stored 
in the envelope at the time $t_{rec}$. As the envelope material recombines,  
it rapidly transforms from opaque to transparent.      
In many Type II supernovae the time until adiabatic expansion cools the 
envelope to the recombination temperature is usually smaller or 
equal to the diffusion time scale and significantly shorter than the 
transparency time scale, so a significant fraction of the initial thermal 
energy is still available at the time of recombination. This energy is then 
released over the time required for the hydrogen envelope to recombine, 
generating what is usually an observable peak in the light curve (SN1987A 
providing a proto-typical example). Roughly, the average luminosity during 
the recombination peak can be estimated as
 \begin{equation} \label{eq:L_rec}
 \bar{L}_{rec}=\frac{E_{thermal}(t_{rec})}{t_{rec}}=
 \frac{4\pi}{3}R_0^3 a \frac{T_{0,H}^2 T_{rec,H}^2}{t_0}\;,
 \end{equation}
assuming that energy losses due to photon diffusion have been negligible, 
which is a good approximation if $t_{rec,H}<t_{diff}$. 
In equation~(\ref{eq:L_rec}) $a$ is the radiation constant. Note that the 
nonlinear nature of the physics governing the propagation of the 
recombination front through the envelope limits the applicability of using 
the average quantities of the envelope to fully describe the light curve 
during the recombination phase 
(see more detailed estimates in Nomoto et al.~1994, Arnett 1996).

\subsection{Radioactive Heating}\label{subsect:basrads} 
 
A complete analysis of a realistic light curve must also include the effects
of heating due to decays of radioactive isotopes synthesized in the
explosion. After recombination, we can estimate that the envelope is
practically transparent to its thermal photons, so that the luminosity due
to radioactive heating is roughly equal to the instantaneous energy
deposition rate due to the decays, through thermalization of the decay
products. The rate of energy deposition in the envelope by the decays of an 
initial total mass $M_X$ of a given isotope $X$ can be expressed as:
\begin{equation}\label{eq:Q_X} 
Q_X(t)=M_X \left( \varepsilon_{X,\gamma}f_X(t)+
\varepsilon_{X,e^+}\right)\mbox{e}^{-t/\tau_X}\;, 
\end{equation} 
where $\tau_X$ is the life-time of the element $X$. The form of 
equation~(\ref{eq:Q_X}) distinguishes between the photons ($\gamma$-rays) and 
positrons emitted in the decays (with energy rates per unit mass of 
$\varepsilon_{X,\gamma}$ and $\varepsilon_{X,e^+}$ respectively) since 
$\gamma$-rays are not totally trapped in the envelope, and 
their contribution to heating is modified by a trapping factor, $f(t)$. On 
average, the hard photons lose about half their energy in every scattering, 
until their energy is degraded to tens of keV.  At these energies
free-bound absorption on heavy elements becomes the dominant source of
opacity and the photons are thermalized rapidly. Because the $\gamma$-ray
opacity rises very rapidly with decreasing energy, we can assume to first
order that once a hard photon scatters it is absorbed. The trapping factor 
can be simply approximated as \citep{Woosleyal89}
\begin{equation}\label{eq:f_X}
f_X(t)=1-\exp\left(-\kappa_{\gamma,X}\Phi_0
\left(\frac{t_0}{t}\right)^2\right)\;,    
\end{equation}
where $\Phi_0(t/t_0)^2$ and $\kappa_{\gamma,X}$ are the $\gamma$-ray 
column depth and opacity
for the typical photons emitted in the decays of element $X$.
Since the inner (more dense) layers are the
dominant contributors to the optical depth for $\gamma$-rays,
the helium-rich layer can be used to estimate the total $\gamma-$ray optical 
depth. The time-dependent column depth is
 \begin{eqnarray}\label{eq:Phi_0}
\Phi_0\left(\frac{t_0}{t}\right)^2 \approx \int \rho dr \approx 
\rho_{He}(t)R_{He}(t) = 
 \rho_{He,0}\left(\frac{t}{t_0}\right)^{-3} V_{He,0}t \\ 
 \;\;\; = \left(\frac{3 M_{He}}{4\pi}\right)^{1/3}\rho_{He,0}^{2/3}
 \left(\frac{t_0}{t}\right)^2\;.  \nonumber
 \end{eqnarray}
In equations~(\ref{eq:f_X}-\ref{eq:Phi_0}) it is implicitly 
assumed that the expansion time $t_0$ is universal to the entire envelope; 
otherwise, it must be replaced with the characteristic expansion time of 
the helium-rich layer. 
Note that equation~(\ref{eq:Phi_0}) allows us to express the trapping factor 
in terms of the initial (and presumably known) global quantities of the 
helium-rich layer. The quantity $\rho_{0,He} t_0^3$ 
(which appears here to the power 
$\frac{2}{3}$) is a useful one also in the context of accretion 
(see Chevalier 1989, and in detail below).

In addition to the characteristic time $\tau_X$, another characteristic 
time involving radioactive heating is 
the $\gamma-$transparency time, similarly to equation~(\ref{eq:t_trns}). 
Since $\gamma$-rays carry the bulk of the energy emitted in the radioactive 
decays, it is important to compare this time, 
\begin{eqnarray}\label{eq:t_trnsgam}
t_{trns-\gamma,X} = 
\left[\frac{3}{4\pi}\kappa_{\gamma,X}M_{He}\right]^{1/2}
\frac{1}{V_{0,He}} \\ 
\;\;\; 
\equiv [\kappa_{\gamma,X} R_{0,He}^3 \rho_{0,He}]^{1/2}{1\over V_{0,He}}\;, 
\nonumber
\end{eqnarray}
with the other characteristic times governing the light curve history. At 
times $t\!\ga\!t_{trns-\gamma,X}$ some of the hard photons emitted in the 
decays of element $X$ escape before thermalizing and do not contribute to 
the bolometric light curve. We note again, that equation~(\ref{eq:t_trnsgam}) 
assumes, as before, that the bulk of the envelope's opacity to hard photons 
is contributed by the helium-rich layer. 
The relevant radioactive isotopes for powering the light curve are \Ni and
its daughter nucleus \Coa, \Cob and \Ti. We adopt the 
characteristic parameters for these isotopes from \citet{Woosleyal89},
listed in table~\ref{tab:radioactives}. Note that in a realistic 
application, especially when the early history of the envelope is of 
interest, the \Coa mass must be adjusted according to its production rate in 
\Ni decays \citep{ShigNom90},
\begin{equation}\label{eq:NiandCo} 
M_{\Coa}(t)=M_{\Ni}(t=0)
\left[\exp(-t/\tau_{\Coa})-\exp(-t/\tau_{\Ni})\right].
\end{equation}

\begin{inlinetable}
\label{tab:radioactives}
\begin{center}
\begin{tabular}{ccccc}
\multicolumn{5}{c}{TABLE 1}\\
\tablevspace{0.2cm}
\multicolumn{5}{c}{The relevant radioactive isotopes for powering a 
supernova light curve:} \\
\multicolumn{5}{c}{
Energy emission rates in $\gamma-$rays and positrons, life-times and}\\ 
\multicolumn{5}{c}{
effective $\gamma-$ray opacities at the characteristic photon energies $^a$}
\\ \tablevspace{0.1cm} 
\tableline
\tableline
\tablevspace{0.1cm}
isotope & $\varepsilon_\gamma$ & $\varepsilon_{e^+}$
& $\tau$ & $\kappa_\gamma/Y_e^T\;\;\;^b$ 
\\
        & (\ergss)             & (\ergss) 
& (days) & (cm$^2$/gm) \\ \tablevspace{0.1cm}
\tableline
\tablevspace{0.1cm}
\Ni  &  $3.90\times 10^{10}$ &        0          & $8.8$             & 0.06 \\
\Coa &  $6.40\times 10^9$    & $2.24\times10^8$  & $111.3$           & 0.06 \\
\Cob &  $6.81\times 10^8$    &  0                & $391.0$           & 0.144\\
\Ti  &  $2.06\times 10^8$    & $6.536\times10^7$ & $3.28\times 10^4$ & 0.073\\
\tableline
\end{tabular}
\end{center}
\end{inlinetable}
$^a${\footnotesize adapted from Woosley et~al.~1989.}\\
$^b${\footnotesize
$Y_e^T =$ the total fraction (bound and free) of electrons per nucleon}

\subsection{Accretion} \label{subsect:Accretion}

Late-time fallback onto the black hole can be characterized in terms of
the ratio of the initial expansion time scale $t_0$ and the initial
{\it accretion time scale},
 \begin{equation}\label{eq:t_acc0}
 t_{acc,0} = \frac{GM_{BH}}{c_{s,0}^3(\mbox{He})}\;,
 \end{equation}
 where $M_{BH}$ is the black hole mass and $c_{s,0}(\mbox{He})$ is the initial
sound speed in the helium rich layer, which we assume is the source of
material for late-time accretion. The hierarchy of these two time scales
determines the hydrodynamic evolution of the accretion flow
\citep{CSW96}. If $\tilde{k}\equiv t_{acc,0}/t_0\gg 1$, the gas has no
time to respond to pressure forces, and accretion proceeds in a dust-like
(pressure-free) manner from its onset. Alternatively, if $\tilde{k}\ll 1$ 
initially, accretion is at first almost unaffected by expansion
and follows a sequence of Bondi-like (Bondi 1952) quasistationary states 
with a slowly decreasing density at large distance. Even in this latter case 
the flow does eventually become dust-like, since expansion continuously
decreases the density and pressure in the reservoir of bound
material. The transition is expected when 
the actual expansion and accretion time scales become comparable, 
yielding a {\it transition time} \citep{CSW96}
 \begin{equation}\label{t_tr}
 \frac{t_{tr}}{t_{acc,0}}\approx
 \left(\frac{9}{2}\right)^{1/(9 \Gamma-11)} 
 \tilde{k}^{-9(\Gamma-1)/(9 \Gamma -
 11)}\;,
 \end{equation}
where $\Gamma$ is the adiabatic index of the gas. 
Thus $t_{tr}\!\approx\!\frac{9}{2} \tilde{k}^{-3}t_{acc,0}$ for
$\Gamma\!=\!4/3$ and $t_{tr}\!\approx
\!\left(\frac{9}{2}\right)^{-1/4}\tilde{k}^{-3/2}t_{acc,0}$ for
$\Gamma\!=\!5/3$.

The hierarchy of time scales reflects the corresponding hierarchy of radii, 
when comparing the marginally bound radius, $R_{mb}$ 
with the accretion radius, $R_{acc}$. Initially
\begin{eqnarray}\label{eq:R_mb-R_acc}
 R_{mb,0}=(2GM_{BH}t_0^2)^{1/3} & , 
& R_{acc,0}=\frac{GM_{BH}}{c_{s,0}^2(\mbox{He})}\;,
\end{eqnarray}
so that if $R_{acc,0}\ll R_{mb,0}$, the initial accretion is Bondi-like. 
As the envelope expands $R_{acc}\sim t$ while $R_{mb}\sim t^{2/3}$ 
\citep{CSW96}, so that, as in the case of time scales, 
eventually the accretion will become dust-like, 
even if it had begun as a fluid one. The transition time $t_{tr}$ roughly 
corresponds to the time when $R_{mb}=R_{acc}$.     

If $t_{acc,0}/t_0 \gg 1$, pressure forces are never dominant and 
the late time accretion rate from an homologously expanding
envelope can be estimated by the dust-like solution of \citet{CSW96}. 
They find that the ratio of the dust-accretion rate $\dot{M}$ to the 
Eddington accretion rate $\dot{M}_{Edd}$ (see below) is 
 \begin{equation}\label{eq:dustmdot}
 \frac{\dot{M}}{\dot{M}_{Edd}} \simeq \frac{4\pi^{2/3}}{9} \rho_{0,He} \,t_0
 \kappa c
 \left ( \frac{t}{t_0} \right )^{-5/3} \, .
 \end{equation}

The resulting luminosity produced by late-time fallback can be computed by 
including radiation in the calculation. 
A self-consistent analysis of the
{\it radiation}-hydrodynamic evolution of the accretion flow from an
expanding envelope \citep{PaperI} shows that, at sufficiently
late times, 
the evolution of the flow always proceeds as a sequence of quasistationary 
states. This is because the dynamical time scale at the accretion or 
marginally bound radius is much longer than all the relevant time scales of 
the flow and the radiation field in the inner accreting region. Both
pressure and radiation forces are negligible at late times, and the flow is
indeed dust-like and accretion rate declines as $t^{-5/3}$. 
If initially $t_{acc,0}/t_0 \ga 1$ and radiation pressure is always negligible
throughout the evolution, $\dot{M}$ is given by 
equation~(\ref{eq:dustmdot}) with the initial characteristic parameters of 
the bound material. On the other hand, if $t_{acc,0}/t_0 \la 1$ and the 
accretion begins as a fluid flow, the late time accretion will still settle 
on a dust-like solution with $\dot{M}\propto t^{-5/3}$, but with a modified 
(reduced) absolute magnitude. 

As a consequence of the quasi-stationary evolution of the flow and the
radiation field, the late-time luminosity produced by fallback from the
expanding envelope closely follows the steady-state, spherical
hypercritical accretion formula derived by Blondin (1986),
 \begin{eqnarray}\label{eq:Blondin_L}
 \frac{L_{acc}}{L_{Edd}} \simeq  4 \times 10^{-7}
 \left( \frac{\mu}{0.5} \right)^{-4/3}
\left( \frac{\kappa}{0.4 \, {\rm cm^2 \, g^{-1}}} \right)^{-1/3} \times  \\
 \;\;\; \left( \frac{M_{BH}}{M_\sun} \right)^{-1/3}
 \left( \frac{\dot M}{\dot M_{Edd}} \right)^{5/6} \; ,   \nonumber
 \end{eqnarray}
for a given accretion rate $\dot{M}$. In equation~(\ref{eq:Blondin_L}) 
$\kappa$ is the opacity and $\mu$ is the mean 
molecular weight per electron of the accreting material. 
The luminosity and accretion rate in equation~(\ref{eq:Blondin_L}) 
are measured in units of the 
appropriate Eddington quantities: the Eddington luminosity, 
$L_{Edd}=4\pi G M_{BH}/\kappa=1.3\!\times\!10^{38}
(\kappa/{\rm 0.4 cm^2 g^{-1}})^{-1}(M_{BH}/M_\sun)\;$ 
\ergss, and  the Eddington accretion rate, 
$\dot M_{Edd}=L_{Edd}/c^2$.
The evolution of the late time accretion luminosity
for a dust-like flow can then be expressed as \citep{SN97Dlet} 
 \begin{equation}\label{eq:L_acc}
 L_{acc}(t) = L_{acc,0}\left(\frac{t}{t_0}\right)^{-25/18}\;,
 \end{equation}
where 
 \begin{equation}\label{eq:L_acc0}
 L_{acc,0}=\Lambda\left(\frac{\mu}{0.5}\right)^{-4/3}
                \left(\frac{\kappa}{0.4}\right)^{-1/2}
 \left(\frac{M_{BH}}{M_\sun}\right)^{2/3}
 \rho_{0,He}^{5/6}t_{0,He}^{20/9}\;,
 \end{equation}
and $\Lambda\approx 1.27\!\times\!10^{40}\;$\ergss~ (for $\rho_{0,He}$ in 
\gmcmc~ and $t_0$ in seconds) arises from the radiative 
efficiency of equation~(\ref{eq:Blondin_L}). {\it It is this power-law decay 
rate in the bolometric luminosity which characterizes the presence of a black 
hole in the aftermath of a supernova explosion}. 

It is important to notice that equations~(\ref{eq:L_acc}) 
and~(\ref{eq:L_acc0}) 
cannot apply at arbitrarily early times even if the initial conditions 
satisfy $t_{acc,0}\ga t_0$. First, some build up time is naturally required 
for the accretion rate
to reach a maximum and start decaying.
This time should be comparable to the shorter of the two time scales 
$t_0$ and $t_{acc,0}$
\footnote{In the case of a pure dust, the build-up time can be estimated by 
comparing the early time accretion rate, which builds up as 
$\dot{M}\propto t$, (Colpi et al.~1996 (eq.~[26])) to the late-time one. 
We find that the two become comparable at a time of $\sim 1.2 t_0$.}. 
Furthermore, if during the early evolutionary stages 
the accretion rate can be so large that the accretion luminosity approaches 
the Eddington limit, radiation pressure will modify the flow even if pressure 
forces were initially negligible. On the basis of
equation (\ref{eq:L_acc}), we can define a critical time
$t_{crit}$ when the luminosity due to dust-like accretion 
equals the Eddington limit
 \begin{eqnarray}\label{eq:t_crit}
 t_{crit}\simeq  14.5\left(\frac{\mu}{0.5}\right)^{-24/25}
                \left(\frac{\kappa}{0.4}\right)^{9/25} 
  \left(\frac{M_{BH}}{M_\sun}\right)^{-6/25} \times  \\ 
  \;\;\; 
 \left(\frac{\rho_{0,He}}{10^{-4}\;\mbox{gm}\;\mbox{cm}^{-3}}\right)^{3/5}
 \left(\frac{t_{0,He}}{1\;\mbox{hr}}\right)^{8/5}\;\mbox{hrs}.  \nonumber
 \end{eqnarray}
According to equation~(\ref{eq:Blondin_L}), the critical rate at which
$L_{acc}\sim L_{Edd}$ is $\dot{M}_{crit}\ga \;1$\msolyr \ for black holes of
several solar masses. If the luminosity reaches the Eddington limit, 
radiation pressure cannot be ignored in comparison to the gravitational
pull of the central object, and so material near the marginally bound
radius (where a significant fraction of the initially bound mass resides)
may receive a sufficient impulse to become unbound. Hence, we expect that
if the build-up time of the flow satisfies 
$\min(t_0,t_{acc,0})\ll t_{crit}$, radiation forces will modify
the accretion history and limit the accretion rate so that 
$\dot{M}(t\ga t_{crit})\approx\dot{M}_{crit}$ by effectively readjusting the
values of $\rho_{0,He}$ and $t_0$ in the bound region.
The total amount of bound material must be
reduced as well. If, on the other hand, 
$\min(t_0,t_{acc,0})\gg t_{crit}$ 
we can expect that radiation pressure will be of
lesser significance throughout the accretion history and that, at late
times, the accretion rate will settle on the dust solution of
equation~(\ref{eq:dustmdot}) with the original values
of $\rho_{0,He}$ and $t_0$ (if $t_{a,0}/t_0 \ga 1$).
 
\subsection{Black Hole Emergence}\label{subsect:emergence}

We define the black hole ``emergence time'', $t_{BH}$ ,  
in a supernova light curve to be the time when the accretion luminosity 
is comparable to all the other sources of luminosity combined, i.e., 
$L_{acc}(t_{BH})=\frac{1}{2}L_{tot}(t_{BH})$. 

Even a small abundance of radioactive isotopes is sufficient to 
impose a bolometric luminosity that will initially dominate accretion 
luminosity, since the latter is inherently limited by the Eddington rate. 
However, continuous spherical accretion must eventually result in emergence, 
since the accretion luminosity (eq.~[\ref{eq:L_acc}]) 
decreases as a power law, while radioactive heating decreases at least 
exponentially, and even more rapidly if $\gamma-$ray trapping is incomplete 
(eq.~[\ref{eq:Q_X}]).
Quantitatively, we expect that for any specific radioactive isotope, at a 
time of several times its life-time, $\tau_X$, the heating rate is declining 
much more rapidly than the accretion luminosity. If the order of magnitude 
of the accretion luminosity and initial abundance of the dominant radioactive 
isotope, $M_X$ are known, the {\it time} of emergence of 
the accretion luminosity is 
determined mostly by the value of $\tau_X$, with some finer tuning 
due to the exact values of $L_{acc}(t)$, $M_X$, and the time dependence of 
$\gamma-$ray trapping. Note that the efficiency of $\gamma-$trapping 
decreases with time as $\e^{(-1/t^2)}$, so the 
dependence of the time of emergence on $M_{X}$ and $L_{acc}$ is even weaker. 
Correspondingly, the {\it luminosity} at emergence is mostly determined by 
$L_{acc}(\mbox{several}\;\tau_X) \propto L_{acc,0}\tau_X^{-25/18}$.

In accordance with \S~\ref{subsect:Accretion}, it is reasonable to assume 
that after a time not longer than a few $\times \min(t_0,t_{acc,0})$ the 
accretion luminosity is modulated to $L_{acc}\la L_{Edd}$.
Hence, for $\min(t_0,t_{acc,0})$ of the order of a few 
days, it is evident that for ``typical'' core collapse supernova, like 
SN1987A, with an inferred $M_{\Coa}\sim 0.1\;M_\sun$ and assumed 
$M_{\Ti}\sim 10^{-4}\;M_\sun$ \citep{WoosWeav95}, the relevant isotope for 
emergence is \Ti, and the time scale for emergence is 
$t_{BH}\approx\mbox{several}\;\tau_{\Ti}\sim$hundreds of years, at which time 
the accretion 
luminosity will drop to a few $10^{32}-10^{33}\;$\ergss. 
Note that since about 
$\frac{1}{3}$ of the decay energy of \Ti is emitted in the form of 
positrons, $\gamma-$ray transparency will not affect these estimates 
significantly. On the other hand, in a supernova where the abundance of 
radioactive elements is reduced by a factor of $\sim35$, as is expected in 
SN1997D (see below), \Coa and \Cob are also relevant for examining emergence. 
The time of emergence is reduced to hundreds of 
days, and the luminosity at emergence will be a few 
$10^{35}-10^{36}\;$\ergss. 
Of course, if no radioactive elements are present, as is expected if the 
mass cutoff in the supernova is very far out, emergence can occur as soon 
as the recombination peak clears the envelope, at $50-100\;$days, with 
$L_{acc}\gtrsim 10^{37}\;$\ergss.

\section{Equations and Numerical Method: Summary and Modifications}
\label{Sect:code}

In order to explore the possibility of detecting black hole
emergence in realistic supernovae in detail, we perform a numerical 
study. Time-dependent simulation of the supernova envelope including 
accretion requires solution of the relativistic radiation hydrodynamic 
equations coupled to the moment equations for radiation transport and 
makes use of a detailed model for the ejecta that includes radioactive 
decay, realistic composition and opacities.
The equations of relativistic, radiation hydrodynamics for a
self-gravitating matter fluid interacting with radiation and the general
relativistic moment equations for the radiation field have been presented
in \citet{PaperI} (to which we refer for details).
These equations were solved using a semi-implicit Lagrangian finite
difference scheme in which the time step is controlled
by the Courant condition and the requirement that the fractional variation
of the variables in one time-step be smaller than ~10\%. 

A fundamental obstacle in tracking the evolution of envelope expansion and 
accretion for several years arises from the extremely wide diversity of 
relevant time scales in the problem. These range from the dynamical time in 
the accreting region (milliseconds close to the 
horizon) to days for the expansion time scale, to hundreds of days for 
the evolution of the light curve. 

Since the innermost layers of the envelope have the shortest physical time 
scales, the computational time can be
considerably reduced with a careful choice of the location of the inner
boundary. During computation, we keep the inner radial boundary of the 
integration domain $R_{in}$ at hundreds to
thousands of Schwarzschild radii, far outside the black hole horizon, 
but still but always {\it well within} the
accreting region.  This measure is acceptable because the gas in the
inner accreting region is in local-thermal-equilibrium (LTE) and in 
near free-fall, thus allowing a
reliable set of boundary conditions at $R_{in}$ \citep{PaperI}.
However, during evolution the effective optical depth in
the accreting region decreases because of the secular decrease in density
caused by expansion. Thus, in order to ensure that the gas is always in
LTE at $R_{in}$, the inner radial boundary must be continuously moved
inward during the simulation. 
We note that with this treatment, the position of the inner boundary usually 
limits the effects of general relativity to be quite small. However, we did 
not alter the general-relativistic nature of the code, especially since  
we cannot determine a priori how far out the inner boundary can be set 
(in particular at late times, when it is continuously being moved inward).  
Furthermore, while the accretion luminosity is insensitive to 
placing the inner boundary away from the black hole horizon for the 
canonical parameters of SN1997D, in more general scenarios it may be 
essential to apply the exact general relativistic formalism with an inner 
boundary in the strong-field domain.

Even taking advantage of the increase in the physical time scales (and
hence in the time step) by suitably positioning $R_{in}$, the 
computational time needed to evolve the solution up to emergence is 
excessive. To reach the emergence stage in a reasonable computational time, 
the code employs a ``Multiple Time-step Procedure'' (MTP), originally 
developed by \cite{PaperI}.  With
this technique, the integration domain is divided into several sub-grids,
and each is evolved on its own characteristic time step. Starting from the
same initial conditions and integration domain, the simulation performed
using this MTP proved to be faster by a factor of $5-8$ than 
one where no time-step acceleration algorithm was employed. 
We note that the most sensitive aspect of the MTP is maintaining accurate 
communication between the subgrids; correspondingly, it can be used safely 
only when there are no sharp features crossing the envelope (i.e., after
recombination).

The main focus of the preliminary investigation of \citet{PaperI} was to
study the fundamental features of fallback in a supernova and, in
particular, to determine the gross properties of the accretion history and
luminosity. To this end, a simplified pure-hydrogen model of the envelope,
generally sufficient to produce a typical early light curve of a Type-II
supernova, was adopted.  Here our goal is to examine black hole emergence
in realistic supernovae, and so we expand the original survey of
\citet{PaperI} to take into account for variable chemical composition of
the expelled envelopes, namely radial dependent abundances of various
elements. In particular, we must account for the presence of radioactive
isotopes and the heating of the envelope by their decays.

\subsection{A Variable Chemical Composition} 

Inclusion of a variable composition is necessary to reproduce realistic
supernova light curves, as has been established in studies of
SN1987A \citep{Woosley88,ShigNom90}. For the purpose of following the
radiation hydrodynamic evolution of the accretion flow onto the black
hole, chemical composition and opacity in the accreting region can have
important quantitative and qualitative effects.  First, there is an
inherent quantitative dependence in the estimate of
equation~(\ref{eq:Blondin_L}), which suggests that a pure hydrogen envelope
produces an accretion luminosity larger by a factor of a few than a
helium-metal-rich one.  Furthermore, it is possible that the presence of
metals in sufficient quantities may impose a qualitative change in the
accretion history due to increased opacity where the material is partially 
ionized. Additional effects
may arise through the the equation of state, but in the case of supersonic
accretion we expect these to be negligible.

In this work we study a supernova envelope with a mixtures of hydrogen, 
helium and oxygen, where the relative partial fractions vary along 
the radial profile of the envelope. 
Oxygen is expected to be the most abundant heavier element in the ejected 
envelope, and we adopt an approximation that it represents the entire metal 
component in the envelope's final composition (we note that over most of 
the density and temperature ranges of interest, an addition of iron at 
solar abundance would change the frequency-integrated opacities by no more 
than $50\%$). For opacity and ionization fractions, we use the TOPS opacities 
available at the LANL server \citep{TOPS}, with extrapolation to low 
temperatures based on \citet{AlexFer94}.
The data are stored in the form of tables, with spacing of five entries per 
decade in both temperature and density for a given composition. 
In the course of computation the Rosseland and Plank opacities and 
ionization fractions are 
calculated with linear interpolation in temperature, density and chemical 
composition. The gas equation of state is 
approximated as an ideal gas of ions and electrons.
     
\subsection{Radioactive Heating}

For the purpose 
of computing the contribution of radioactive decays to the light curve, 
a consistent treatment of radioactive heating is required to follow 
the trapping efficiency, $f_\gamma$, throughout the envelope. If the rate
of radioactive heating is large enough it may also affect the 
hydrodynamic evolution of the envelope material (there exists observational 
evidence for such an effect by \Ni decay in the early history of SN1987A 
- the ``Ni-bubble'', Arnett et al.~1989). The role of radioactive heating 
may be especially important in the case of interest here, since most of the 
radioactive elements are likely to lie in the inner part of the envelope, 
which is the reservoir for accretion. The exact distribution will depend on 
the extent of mixing caused in the process of the explosion 
\citep{ShigNom90,Bethe90}.
Significant heating at early times may accelerate some of the material which 
was initially bound to the black hole and unbind it, hence imposing a 
smaller late-time accretion rate. On the other hand, at late times, even a 
small amount of heating can maintain a finite ionization fraction, 
which in turn will impose a larger photospheric radius, hence
affecting the luminosity.

In our numerical study we introduce an effective treatment of 
radioactive elements by including a local heating rate. We assume that 
the abundance of radioactive elements is proportional to the oxygen
fraction \citep{PinWoos89}, with the relevant radioactive isotopes listed in 
table~\ref{tab:radioactives}. We use the approximation discussed in 
\S~\ref{subsect:basrads} that a $\gamma$-ray photon is absorbed after its 
first scattering.
The trapping factor of the $\gamma$-rays emitted 
at radius $r$ in the envelope can then be estimated as
\begin{equation}\label{eq:f_gam}
f_{X,\gamma}(r)=
1-exp\left\{-(\int_r^{R_{out}} \kappa_{X,\gamma} \rho(r) d\!r)\right\}\;,
\end{equation}
where the quantity in parenthesis is the appropriate $\gamma$-optical
depth at 
radius $r$. In our simulation we impose a further approximation that 
all the trapped energy is absorbed locally at the point of emission. We note 
that this simplified treatment of radioactive decay energy deposition is
satisfactory during the first evolutionary stages when the ejecta are
still very optically thick to $\gamma$-rays. Later on, however, the
constant opacity/single scattering assumption gives only an approximate
description of the transfer of $\gamma$-rays through the envelope. In
fact, as the ejecta become marginally thick, photons can be scattered or
absorbed after traveling a significant distance, making non-local effects
important. An accurate calculation of the energy deposition and the
$\gamma$-ray luminosity would require the solution of the energy-dependent
transport problem for the $\gamma$ rays, which we do not attempt.
Our effective approach reduces the treatment of 
radioactive heating to an additional local energy source. By determining 
the total deposited energy rate
\begin{equation}\label{eq:Q}
 Q = \sum_X Q_X\;,
\end{equation}   
through equations~(\ref{eq:Q_X}) and~(\ref{eq:f_gam}), we modify the energy
equation (eq.[25] of Zampieri et al.~1998a) by 
adding the energy input from radioactive decay
\begin{equation}\label{eq:ener}
\epsilon_{,t} + \widetilde{a} k_P (B-w_0) +p \left(\frac{1}{\rho}\right)_{,t}
= Q \;,
\end{equation}
where $P$, $\rho$ and $\epsilon$, are the pressure, mass density and 
internal energy per unit mass of the gas, $\widetilde{a}$ is related to
the $00$ coefficient of the spherically symmetric, comoving-frame metric,
$\kappa_p$ is the Plank mean opacity, $B\equiv a T^4$ and $4\pi w_0$ is
the radiation energy density (see Zampieri et al.~1998a for details).
All these quantities are measured in the comoving frame.

With the aid of this modified numerical code and our analytic estimates 
we now proceed to study the evolution 
and light curve of realistic supernovae including fallback onto a nascent 
black hole. Our emphasis in this work is the peculiar Type II 
SN1997D, that may provide the first direct observational 
evidence of black hole formation in supernovae. 
This source is the only current 
candidate where an actual observation of the light curve at emergence of the 
black hole may be (marginally) possible.

\section{Black Hole Emergence in SN1997D}\label{Sect:SN1997D}
\vspace{0.15cm}

SN1997D was serendipitously discovered on January 14, 1997 by
\citet{deMello97} during an observation of the parent galaxy NGC 
1536. In their analysis of the bolometric light curve and spectra, 
\citet{Turatto98} found that SN1997D was an extremely peculiar Type II 
supernova. Modeling of the early light curve indicates that the supernova 
was discovered close to maximum light (hence, presumably, at the
recombination peak). 
Adopting a distance modulus of $\mu\!=\!30.64$ (a distance of $\sim\!14\;$Mpc 
to NGC 1536), implied that the maximum absolute magnitude of SN1997D was 
$M_V\!\ga\!-14.65\; (L_{max}\la10^{41}\;$\ergss), fainter by about 2 
magnitudes than typical Type II supernovae at maximum \citep{Patat94}: 
SN1997D is possibly the most sub-luminous Type II supernova ever observed.

\subsection{Fundamental Estimates about the character of SN1997D}  

The most unique feature of SN1997D in the context of black hole emergence
is the very low mass of \Coa deduced from the light curve tail (beginning
at $\sim\!30\;$days after the first observation). The magnitude and shape
of the tail are consistent with a very low abundance of \Coa,
$M_{\Coa}\!\approx\!2\!\times\!10^{-3}\;M_\sun$ \citep{Turatto98}. This
important result has been recently confirmed by \citet{Benetti00} 
on the basis of a new set of photometric data which extends up to 1.5 years 
after the explosion.
  
\citet{Turatto98} found that the light curve and spectra were best 
reproduced with $4\!\times\!10^{50}\;$ergs explosion of a $26\;M_\sun$ main 
sequence star (see Young et al.~1998 for details). The progenitor structure 
prior to collapse was an $8\;M_\sun$ helium-rich core surrounded by an
$18\;M_\sun$ hydrogen-rich envelope. The remnant formed in the supernova 
is expected to have a total mass of about $3\;M_\sun$, 
composed of a $1.8\;M_\sun$ collapsed core and about $1.2\;M_\sun$ of 
material which was accreted in early fallback 
(T.~R.~Young 1999, private communication). The large amount of fallback is 
a consequence of the low explosion energy, and provides a natural 
explanation for the very low observed \Coa mass.
Since $3\;M_\sun$ is almost certainly larger than the maximum mass of 
neutron stars (even a rapidly rotating one; Cook, Shapiro \& Teukolsky 1994), 
we can expect that a black hole was formed by SN1997D.

We note that a different model of a low mass ($\sim\!8\;M_\sun$)
progenitor was suggested by \citet{ChugUt99} to 
explain the early light curve, as well as a low \Coa abundance.
In such an explosion little fallback would have occurred and the remnant
would be most likely a neutron star. The late time light curve shows no
evidence of the presence of a neutron star \citep{Benetti00}, which would
have been observable if it were a Crab-like pulsar, or if it was
spherically accreting from late-time fallback. Hence a black hole 
(and therefore a massive progenitor) is more consistent with the existing 
photometric data.

Assuming that the black hole model for SN1997D is correct, its remnant 
includes a black hole surrounded by a low-velocity ejecta with a 
small abundance of radioactive elements. If late-time fallback is 
proceeding in spherical accretion, the black hole could emerge rather early, 
while it is still potentially detectable. With a preliminary analytic 
estimate, \citet{SN97Dlet} found that emergence could occur as early as 
three years after the explosion and that the accretion luminosity at the 
time is roughly in the range $10^{35}-5\!\times\! 10^{36}$\ergss. 

We have reconsidered black hole emergence in SN1997D 
to determine if the accretion tail persists in a realistic
model of the ejecta that includes a variable composition, realistic
opacities and radioactive heating and takes into account the radiation
hydrodynamic evolution of the helium mantle starting from the conditions
at few tens of hours after the explosion.  In the following we 
attempt to constrain the luminosity at the epoch of emergence, 
using both analytic estimates and numerical simulations.

\subsection{Analytic Estimates for SN1997D \label{subsect:97Danalytic}}

In an analytic approach, we can use the average quantities of the 
helium-rich layer to estimate the radioactive heating and accretion 
luminosities, since this layer is expected to be both the source of material 
for late time fallback and the main contributor to $\gamma-$ray trapping.
As a reference composition of the bound material we 
use the best fit model of \citet{Young98} (see below) which is -- 
by mass -- H:0.10, He:0.45, O:0.45. The electron scattering opacity
at full ionization is $\kappa_{es}=0.22$ and the atomic weight per electron 
is $\mu\approx1.27$.  

\vspace{0.3cm}
\subsubsection{Initial Density and Expansion Time Scale 
\label{ssubsect:anarho0t0}}

Note that for initially homologous expansion, the total kinetic energy of 
the layer can be related to its total mass, $M$, and radius, $R$, according to
\begin{equation}\label{eq:Ekinhom}
E_{kin}=\frac{3}{10}M V^2=\frac{3}{10}M \left(\frac{R}{t_0}\right)^2,
\end{equation}
where $V$ is the velocity at the outer radius and $t_0\equiv R/V$ is the 
corresponding expansion time scale. By expressing the radius in terms of 
the mass and initial average density, $\rho_0$ we can write
\begin{equation}\label{eq:MEandrhot3}
(\rho_0 t_0^3)=\left(\frac{3}{10}\right)^{3/2}\frac{3}{4\pi}
M\left(\frac{M}{E_{kin}}\right)^{3/2}\;.
\end{equation}

Equation~(\ref{eq:Ekinhom}) is especially convenient, since the 
average energy per unit mass deposited in a roughly homogeneous layer is 
relatively easily determined in simulations (see Woosley 1988 for 
SN1987A). The fraction of the total explosion energy carried as kinetic 
energy of the helium-rich layer is typically a few percent. Adopting the 
global quantities of the model of \citet{Turatto98}, we have in the case of 
of SN1997D a helium-rich layer with $M_{He}\approx 5\;M_\sun$, 
$E_{kin,He}\approx 1.2\!\times\!10^{49}\;$ergs, or 
$E_{kin,He}/M_{He}\approx 1.2\!\times\!10^{15}\;$\ergsgm. 
For this layer 
\begin{equation} \label{eq:rho0t03is}
\rho_0 t_0^3\approx9.4\times 10^{9}\mbox{gm cm}^{-3}\;\mbox{s}^3\;.
\end{equation}
Note that the relatively low energy of the explosion is clearly reflected in 
this quantity, which for SN1987A is only 
$\sim\!10^9 \mbox{gm cm}^{-3}\mbox{s}^3$ \citep{Chevalier89}.

The value of $\rho_0 t_0^3$ is sufficient for estimating $\gamma-$ray 
trapping (eq.~[\ref{eq:t_trnsgam}]), but additional information is required 
for estimating the accretion luminosity, which scales as 
$(\rho_0 t_0^3)^{5/6}t_0^{-1/4}$ (eq.~[\ref{eq:L_acc}]). This difference is 
an inevitable consequence of the fact that the bound material is not actually 
expanding homologously, since the gravitational deceleration cannot 
be neglected. For a realistic supernova envelope there is no 
time instant when the entire envelope is in coherent homologous motion. As 
a first approximation, we assume that the entire envelope has a common 
expansion time, which turns out to be $t_0\approx 30\;$hrs for an initial 
radius of $2\!\times\! 10^{13}$ and a hydrogen envelope of $18\;M_\sun$. 
We also have $\rho_{He,0}\approx 7.45\!\times\!10^{-6}\;$\gmcmc~ for these 
parameters. The total mass bound to a black hole of $3\;M_\sun$ is then
\begin{equation}\label{eq:Mbound}
M_{bnd}=\frac{4\pi}{3}\rho_{He,0}R_{mb,0}^3=\frac{8\pi}{3}G M_{BH} t_{0}^2 
\rho_{0,He}=0.136\;M_\sun\;,
\end{equation}
where we have used equation~(\ref{eq:R_mb-R_acc}).

\vspace{0.15cm}
\subsubsection{Accretion Time Scale \label{anatacc}}

Homologous expansion implies that the thermal energy is sufficiently 
smaller than the kinetic energy. In particular, in order to have 
$E_{th}/E_{kin}\la 0.1$ for the helium-rich layer presented above, 
the temperature in the layer must be $T_{0,He}\la 8\!\times\!10^5\;$\Kelv. 
The corresponding gas and radiation pressure are of similar magnitude, 
leading to an estimate of the initial sound speed in this layer of 
$c_{s,0}(\mbox{He})\approx 10^7\;$\cms. 
The initial accretion time is thus $t_{acc,0}\approx120\;$hours, 
significantly longer than the expansion time scale. In this analytic    
estimate, the accretion flow is approximately dust-like from its onset.
As long as we assume that the total thermal energy is the order of 
$\sim10\%$ of the kinetic energy of the helium layer, which itself is 
well constrained, the temperature in this layer is approximately fixed at 
the above value. The initial gas pressure scales as $P_0\sim T\!\rho_0$, 
so the sound speed cannot decrease significantly below 
$c_{s,0}(\mbox{He})\approx 10^7\;$\cms (if the pressure is gas dominated 
$c_{s,0}^2(\mbox{He})\propto P_0/\rho_0\propto T$), 
and the initial accretion time 
scale (\ref{eq:t_acc0}) cannot deviate sharply from 
$t_{acc,0}\approx 100\;$hrs, 
even if the combination of $\rho_0$ and $t_0$ is allowed to vary.    

\vspace{0.15cm}
\subsubsection{Critical Time \label{anatcrit}}

A key feature in this analytic estimate is that 
$t_{crit}\approx 180\;$hrs (eq.~[\ref{eq:t_crit}]) 
or $\sim 6 t_0$. As discussed in \S~\ref{Sect:basics}, dust-like accretion
should require a time of about $t_0$ to settle into the late-time 
temporal decay. Hence it is likely that radiation forces will impose 
some modification to the accretion flow. We can expect that when the 
accretion flow finally resettles on a dust-like motion the total bound mass 
will be reduced, and the average density and expansion time in the bound 
region will be modified. The initial values of $\rho_0$ and $t_0$ can 
thus provide an upper limit on the late time accretion rate and luminosity 
through equations~(\ref{eq:dustmdot}) and ~(\ref{eq:L_acc}). 

The effect of radiation pressure on the early-time 
accretion flow when the luminosity is close to the Eddington limit is 
non-linear, and therefore cannot be ascertained from our analytic 
treatment. However, by making two simplifying assumptions we can 
roughly estimate the impact of the Eddington-limited stage on the late-time 
accretion rate. First, we assume that the accretion flow adjusts so that 
the accretion luminosity is exactly the Eddington value. Effective 
gravity is then negligible everywhere. Thus, while the average density 
$\widetilde{\rho}_0(t)$ and expansion time $\widetilde{t}_0(t)$ 
maintain a constant $\widetilde{\rho}_0(t)\widetilde{t}_0(t)^3=\rho_0 t_0^3$, 
individually they vary according to
\begin{equation}\label{eq:rho0-t_t0-t}
\widetilde{\rho}_0(t)=\left(\frac{R_0+V_0 t}{R_0}\right)^{-3}\;,\;
\widetilde{t}_0(t)=t_0\left(\frac{R_0+V_0 t}{R_0}\right)\;.
\end{equation}
The combination $\rho_0 t_0^{8/3}$, which determines the magnitude of 
the dust-like accretion rate (eq.~[\ref{eq:dustmdot}]) evolves as 
\begin{equation}\label{eq:rho0t083}
\widetilde{\rho}_0(t)\widetilde{t}_0^{8/3}(t)
=\rho_0 t_0^{8/3}\left(1+\frac{t}{t_0}\right)^{-1/3}\;.
\end{equation}
The second assumption is that the transition to a dust-like flow is 
instantaneous, and that at the time of transition, $t_{dust}$, the values 
$\widetilde{\rho}_0(t_{dust})$ and $\widetilde{t}_0(t_{dust})$ exactly
yield $\dot{M}(t_{dust})=\dot{M}_{crit}$ through 
equation~(\ref{eq:dustmdot}). From equation~(\ref{eq:rho0-t_t0-t}) 
we find that $t_{dust}$ must then satisfy the relation:
\begin{eqnarray}\label{eq:t_dust}
 \!\!\! 41.15\left(\frac{\mu}{0.5}\right)^{-4/3}
                \left(\frac{\kappa}{0.4}\right)^{1/2}
 \left(\frac{M_{BH}}{M_\sun}\right)^{-1/3} \nonumber \\ \;\;\;\ 
 \times 
 \left(\frac{\rho_{0,He}}{10^{-4}\;\mbox{gm}\;\mbox{cm}^{-3}}\right)^{5/6}
 \left(\frac{t_{0,He}}{1\;\mbox{hr}}\right)^{5/6}  \\
 \;\;\;\;\;\; \times \left(1+\frac{t_{dust}}{t_0}\right)^{-5/18} 
 \left(\frac{t_{dust}}{t_0}\right)^{-25/18}=1 \;.  \nonumber
\end{eqnarray}

The modification due to radiation pressure is that $t_{dust}<t_{crit}$, 
so that the late-time accretion luminosity is
$L(t)=L_{Edd}(t/t_{dust})^{-25/18}$ instead of 
$L(t)=L_{Edd}(t/t_{crit})^{-25/18}$. We find that 
$t_{dust}\approx 129\;$hrs, so the late time
accretion luminosity will be reduced by a factor of $\sim0.633$ compared
to that predicted by equation~(\ref{eq:L_acc}) for the initial parameters. 

Note that in the limit $t_{dust}\gg t_0$ equation~(\ref{eq:t_dust})  reduces 
to the form $t^2_{dust}\propto \rho_0 t^3_0$, so $t_{dust}$ is 
dependent only on the initial specific energy of the bound material. This 
is a natural consequence of the 
approximation used here, since ongoing homologous expansion eventually 
loses memory of the initial size $R_0$ (which determines $t_0$ and 
$\rho_0$ separately).
For the values presented above, $t_{dust}$ is 
larger than $t_0$ by only a factor of a few so the above argument is not 
exact, but if $t_0$ is confined to a reasonably small range, 
the late-time accretion luminosity can still be very well constrained by 
the specific energy of the helium-rich layer. 
Indeed, we find that $t_{dust}$ varies by less than $10\%$ when $t_0$ 
is varied in the range $20-40\;$hrs if $\rho_{0,He} t^3_0$ is kept fixed.

\vspace{0.3cm}
\subsubsection{Radioactive Heating and an Estimate of Emergence}
\label{anaemrg}

\begin{deluxetable}{c c c c c c c c c c}
\tablefontsize{\small}
\tablewidth{0pt}
\tablecaption{Estimates of black hole emergence in SN1997D 
based on the analytic model (see text). 
\label{tab:ResAnalytic}} 
\tablehead{ 
\colhead{$E_{He}/E_{tot}$} & 
\colhead{$\rho_0 t_0^3$} & 
\colhead{$t_0$} &
\colhead{$\rho_0$} &
\colhead{$t_{crit}$} &
\colhead{$t_{dust}$} &
\colhead{$t_{BH}$\tablenotemark{a}} &
\colhead{$L_{tot}(t_{BH})$\tablenotemark{a}} &
\colhead{$t_{BH}$\tablenotemark{b}} &
\colhead{$L_{tot}(t_{BH})$\tablenotemark{b}}  \\
\colhead{} &
\colhead{($10^9\;$\gmcmc s)} &
\colhead{(hrs)} &
\colhead{($10^{-5}$\gmcmc)} &
\colhead{(hrs)} &
\colhead{(hrs)} &
\colhead{(days)} &
\colhead{($10^{36}\;$\ergss)} &
\colhead{(days)}   &
\colhead{($10^{36}\;$\ergss)} 
}
\startdata
0.01 &  48.75 &  20  & 13.11 & $\sim524$ & $\sim302$ 
     &  1148  & 2.69 & 1090  & 2.91 \\
0.01 &  48.75 &  30  & 3.87  & $\sim483$ & $\sim301$ 
     &   1149 & 2.68 & 1092  & 2.88 \\
0.01 &  48.75 &  40  & 1.63  & $\sim456$ & $\sim299$ 
     &   1151 & 2.65 & 1095  & 2.85 \\
0.03 &   9.38 &  20  & 2.51  & $\sim195$ & $\sim131$  
     &   1455 & 0.61 & 1218  & 0.78 \\ 
0.03 &   9.38 &   30 & 0.75  & $\sim180$ & $\sim129$ 
     &   1466 & 0.59 & 1216  & 0.72 \\
0.03 &   9.38 &   40 & 0.31  & $\sim170$ & $\sim128$  
     &   1472 & 0.58 & 1220  & 0.75 \\
0.05 &   4.36 & 20 &  1.17 & $\sim123$ & $\sim88$ & 
   $\sim1500$\tablenotemark{\dagger} &
   $\sim0.40$\tablenotemark{\dagger} & 1254 & 0.43\\
0.05 &   4.36 & 30 &  0.35 & $\sim114$ & $\sim87$ & 
   $\sim1500$\tablenotemark{\dagger} & 
   $\sim0.39$\tablenotemark{\dagger} & 1256 & 0.42 \\
0.05 &   4.36 & 40 &  0.15 & $\sim107$ & $\sim86$  & 
   $\sim1500$\tablenotemark{\dagger} & 
   $\sim0.38$\tablenotemark{\dagger} & 1260 & 0.41 \\
\enddata
\tablenotetext{a}{
\Cob and \Ti abundances scaled according to SN1987A}
\tablenotetext{b}{
No \Cob and \Ti (only \Coa)}
\tablenotetext{\dagger}{
In this model the accretion luminosity does not reach $50\%$ of the total 
luminosity (although it does constitute more than $40\%$)}
\end{deluxetable}

With all the necessary values for estimating the late-time accretion rate 
and $\gamma$-ray opacities, we now proceed to examine the competition between 
the accretion and radioactive luminosities. 
The exact time and luminosity at emergence depend 
on the abundances of \Cob and \Ti in the envelope. 
These two isotopes are mostly produced in deeper layers than those
where \Coa is synthesized (especially \Ti, which is produced through 
Nuclear-Statistical-Equilibrium (NSE) in the innermost layers 
of the shocked envelope material, Timmes et al.~1996). We examine their 
effects with two extreme assumptions -- either that they are 
completely absent, or that their abundances scale with that of 
\Coa. In the latter case, using 
$M_{\Cob}/M_{\Coa}(\mbox{SN1987A})=0.0243,\; 
M_{\Ti}/M_{\Coa}(\mbox{SN1987A})=1.33\!\times\!10^{-3}$ we obtain
$M_{\Cob}(\mbox{SN1997D})\la5\!\times\!10^{-5}\;M_\sun,\;
M_{\Ti}(\mbox{SN1997D})\!\la\!2.7\!\times\!10^{-6}\;M_\sun$. 
Assuming that the opacity to $\gamma-$rays is independent of the accretion 
history (as is reasonable, since the initially bound region is a small 
fraction of the total helium-rich layer), we can treat the radioactive 
heating history as fixed once the initial combination $\rho_{0,He} t^3_0$ is 
given (eq.~[\ref{eq:Phi_0}]).

The resulting luminosities due to accretion and radioactive heating 
for our reference 
parameters of $t_0=30\;$hrs and $\rho_{0,He}=7.45\!\times\!10^{-6}\;$\gmcmc 
are shown in Fig.~\ref{fig:analytics}.
If the abundances of \Cob and \Ti are negligible, the total luminosity due to 
radioactive heating is equal to the heating rate through decays of \Coa. 
In this case, the accretion luminosity does emerge with 
$L_{acc}(t_{BH})=L_{\Coa}(t_{BH})\approx 3.6\!\times\! 10^{35}\;$\ergss at 
a time of $t_{BH}\approx1216\;$days (April 2000), after which the 
light curve will settle on a power law decline rather quickly.
If \Cob and \Ti are present with abundances rescaled from SN1987A their 
impact on the luminosity at $\sim 1000\;$days is not negligible, and 
accretion luminosity must be compared to the total radioactive heating 
(labeled $L_{tot-rad}$ in Fig.~\ref{fig:analytics}). In this case 
emergence is delayed to $t_{BH}\approx1466\;$days (December 2000), 
with $L_{acc}(t_{BH})=L_{tot-rad}(t_{BH})\approx 
2.9\!\times\! 10^{35}\;$\ergss. 
Note that in this case the contribution of \Ti
is about $1.4\!\times\!10^{35}\;$\ergss, which decreases 
only very slowly due to increasing $\gamma-$ray transparency. 
Correspondingly, the total luminosity (labeled $L_{tot}$) does not follow a 
perfect power law until \Ti heating becomes negligible, 
although the accretion luminosity will remain the dominant 
source of luminosity for several thousands of days after $t_{BH}$.

The results of similar calculations for the emergence of a $3\;M_\sun$ 
black hole in SN1997D for different values of the initial parameters
are presented in table~\ref{tab:ResAnalytic}. 
We vary the energy of the $5\;M_\sun$ helium-rich layer as 1, 3 
(the nominal case) and 5 percent of the total explosion energy of 
$4\!\times\! 10^{50}\;$ergs, while varying $t_0$ in the range $20-40\;$hrs. 
We list the derived values of $t_{crit}$ and $t_{dust}$ for each model and 
the resulting luminosity and time of emergence. Due to uncertainty regarding 
the abundances of \Coa and \Ti, we examine the two extreme cases where their 
abundances scale according to SN1987A, and when they are completely absent.  
The case of the helium-rich layer having $5\%$ of the total energy and with 
the significant abundance of \Cob and \Ti marks the border-line of 
achieving emergence after a few years (instead of in thousands of years) 
as the the combination of \Cob and \Ti heating is slightly {\it larger} 
than the accretion luminosity (by about $\sim20\%$) for a very long time.

Our results indicate that if a black hole was formed, it is likely to emerge 
about three years after the explosion. Taking into account also for 
the uncertainties in composition and black hole mass, we estimate that 
plausible values of the luminosity 
at emergence seems to lie in the range $0.3-3\!\times\!10^{36}\;$\ergss, 
and the expected time of emergence is $\sim1100-1500$ days after the 
explosion, with higher luminosities corresponding to earlier times. We 
emphasize that the dominant quantity that determines the value of the 
luminosity at emergence is the kinetic energy of the helium-rich layer, while 
the time of emergence and the character of the late-time light curve are also 
sensitive to the possible presence of \Cob and \Ti in 
non-negligible abundances. 

\vspace{0.3cm}
\begin{inlinefigure}
\centerline{\includegraphics[width=1.0\linewidth]{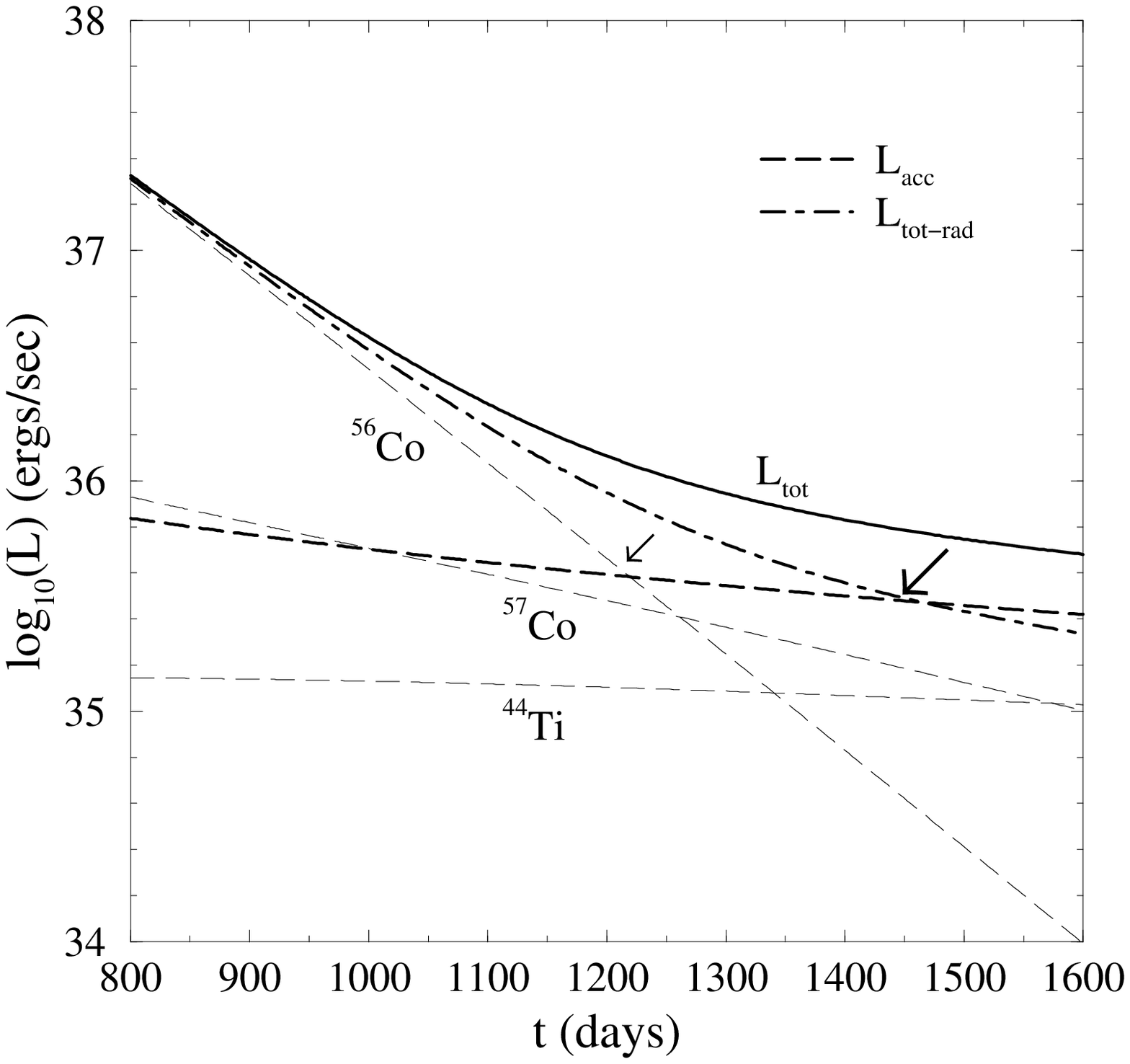}}
\figcaption{Analytic estimates for the total 
bolometric luminosity of SN1997D, $L_{tot}$,  
and the partial contributions of accretion onto the black hole ($L_{acc}$) 
and of radioactive heating 
($L_{tot-rad}=L_{^{56}Co}+L_{^{57}Co}+L_{^{44}Ti}$) 
during the period $800-1600\;$days from the explosion (Jan.~1999-Mar.~2001). 
Arrows mark the time of emergence of the accretion luminosity in the case 
where (1) large arrow: \Cob and \Ti are assumed to be present in the 
envelope with abundances scaled down from SN1987A (see text) and 
(2) small arrow: when only \Coa is present.
\label{fig:analytics}}
\end{inlinefigure}

\vspace{0.15cm}
\subsection{Numerical Results for SN1997D}

While an approximate analytic treatment and scaling behavior 
can provide useful estimates of the time and luminosity at emergence,
it also underscores the need to perform numerical
investigation which can track in detail the early time accretion history
and the late-time evolution of the light curve. The numerical simulation 
is also required to quantitatively assess the role of non-linear 
radiation-hydrodynamics processes and the effects of variable chemical 
composition, realistic opacities and radioactive heating. 
In order to explore in detail the emergence of a black hole in SN1997D, a
full radiation-hydrodynamic simulation of fallback of material from the
supernova envelope up to the several years after the explosion has been
performed, using the numerical code described in \S~3. The initial
conditions for the ejecta were based on the best-fitted model of
\citet{Young98} for the SN1997D explosion. This model of a $26\;M_\sun$ 
progenitor was fitted on the
base of the early time light curve (up to $\sim 100\;$days after the 
explosion), which is practically independent of
the inner part of the helium-rich layer. Hence, while a simulation based on 
this model serves as an essential quantitative estimate, the unavoidable 
uncertainties regarding the initial velocity, density and temperature 
profiles of this inner layer obviously limit the applicability of the 
simulation as a ``best fit'' also for the accretion history and luminosity.

\vspace{0.15cm}
\subsubsection{Initial Profile}

The initial profile describes the supernova envelope at the time 
of break-out of the shock at the surface, corresponding to about 
13 hours after core bounce. The expanding envelope is composed of two main
components: an inner helium-rich layer and an outer hydrogen-rich
envelope. The properties of the profile are presented in
Figs.~\ref{fig:init} and \ref{fig:composition} and the region of initially
bound material is shown in more detail in Fig.~\ref{fig:initbound}. 
By assuming that the abundance of radioactive isotopes scales with the 
oxygen mass fraction, we can simply estimate the mass fraction of \Ni in 
every mass shell by requiring that the total \Ni mass is 
$2\!\times\!10^{-3}\;M_\sun$ (equal to the total \Coa mass inferred by 
observations). In the simulation we assume that the abundances 
of \Cob and \Ti scale with the abundance of \Ni, so their total masses are 
set according to the aforementioned quantities.

In the context of late-time accretion, the most dominant feature of the 
initial profile is the non-uniformity of the helium-rich layer. 
A significant fraction of the mass is 
concentrated in an over-dense region bordering with the hydrogen-rich layer. 
This region also holds the bulk of the kinetic energy of the helium-rich 
layer, $1.7\!\times\!10^{15}\;$\ergsgm, somewhat larger than the 
average value of $1.2\!\times\!10^{15}\;$\ergsgm~ mentioned above. 
The bound region on the other hand, is relatively under-dense and slow. 
The total initially bound mass is $\sim0.22\;M_\sun$.

The expansion time scale throughout most of the initially bound material is  
$t_0\approx 50-60\;$hrs. This region is radiation-pressure dominated, and 
the initial sound speed is roughly 
$c_{s,0}(\mbox{He})\approx 1.3\!\times\!10^7\;$\cms. 
The corresponding accretion time scale, $t_{acc,0}\!\approx\!50\;$hrs 
is therefore similar to the initial expansion time scale. Unlike the analytic 
cases discussed above, in this model pressure forces {\it are} initially 
important. The long initial expansion time scale more than 
compensates for the decreased density in terms of determining the accretion 
rate, and we have $t_{crit}\simeq 300\;$hrs. This model is also 
expected to reach very high accretion rates at early times, which will  
result in some moderation of the accretion flow  
when $\dot{M}_{crit}\approx3.13\;$\msolyr~ is approached.

\begin{inlinefigure}
\centerline{\includegraphics[width=1.0\linewidth]{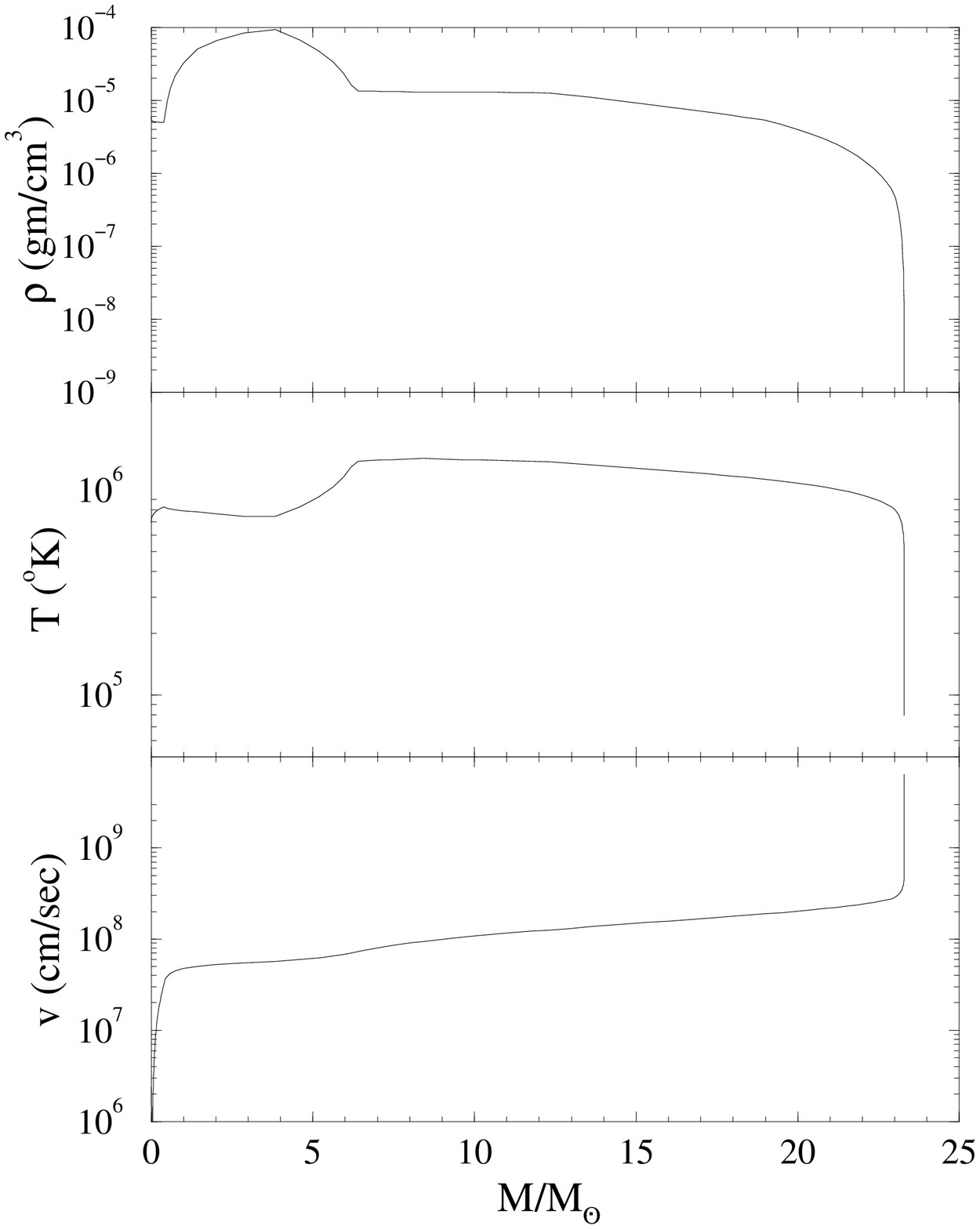}}
\figcaption{Initial SN1997D profile for the fiducial model of 
\citet{Young98}: the density, temperature and velocity profiles of the 
envelope at shock emergence ($\sim 13\;$hrs after core bounce).
\label{fig:init}}
\end{inlinefigure}

\begin{inlinefigure}
\centerline{\includegraphics[width=1.0\linewidth]{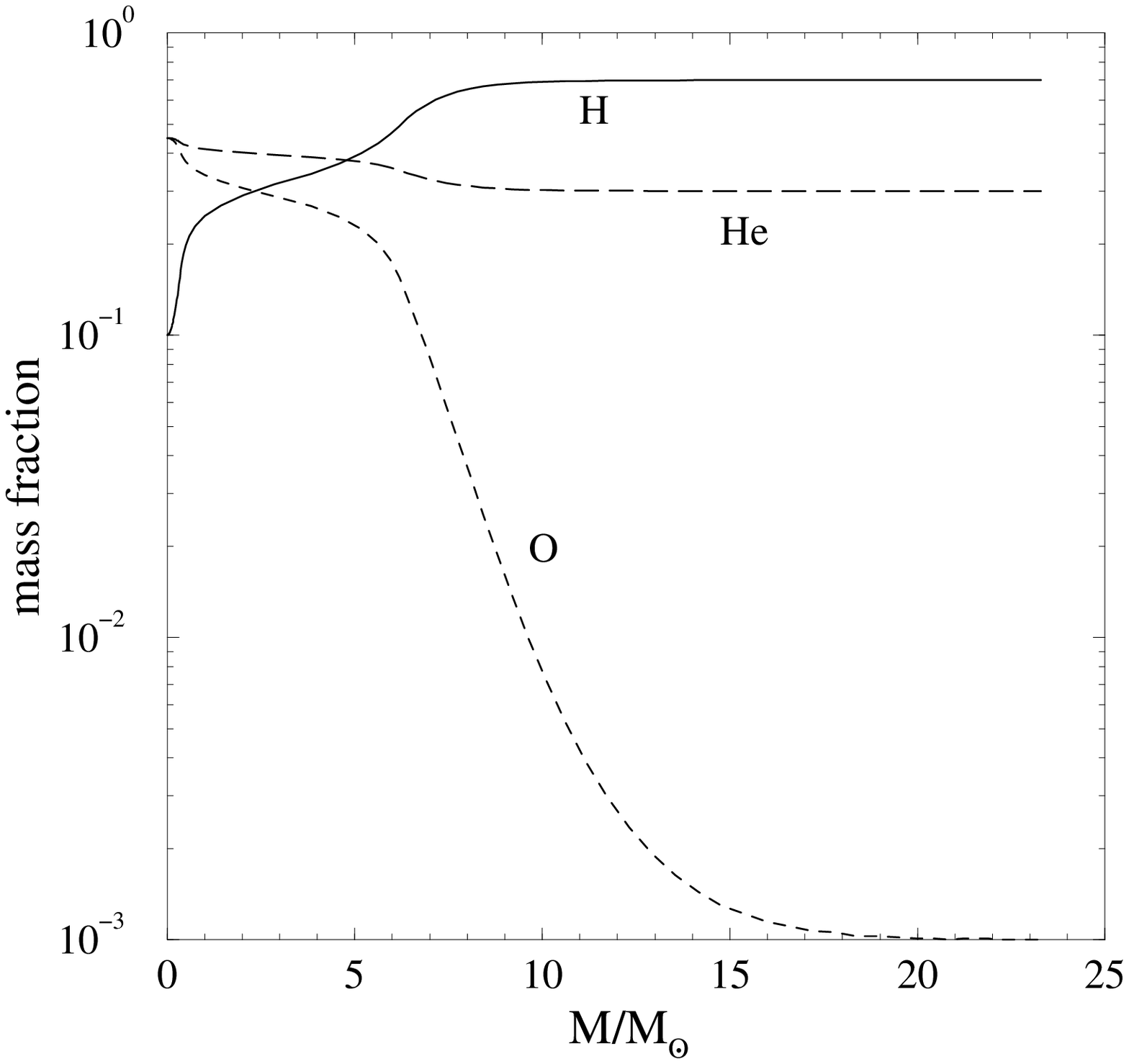}}
\figcaption{Initial SN1997D profile for the fiducial model of 
\citet{Young98}: the composition profile (mass fractions of H-He-O) for the 
envelope.
\label{fig:composition}}
\end{inlinefigure}

\begin{inlinefigure}
\centerline{\includegraphics[width=1.0\linewidth]{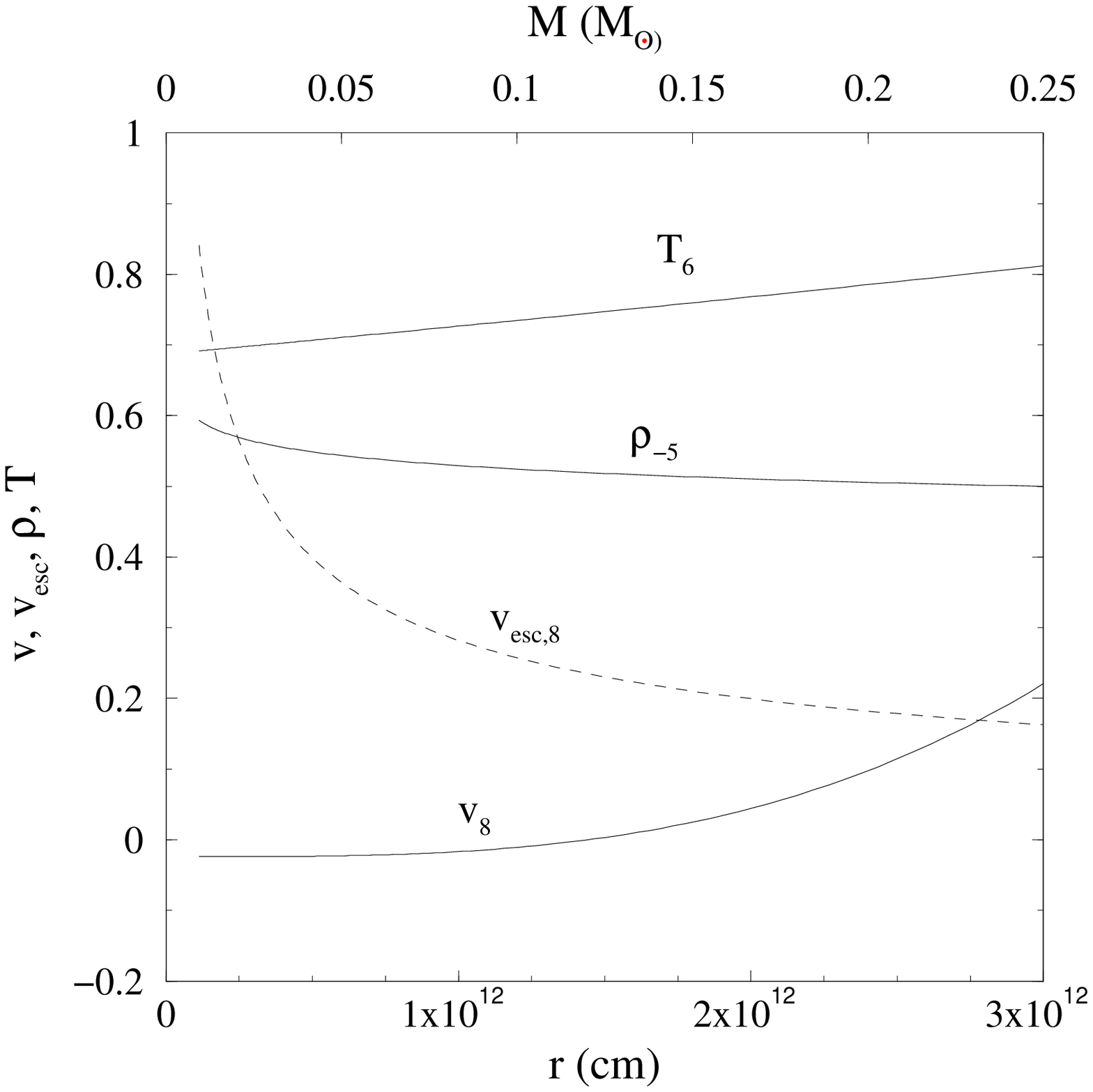}}
\figcaption{Blow up of the initial SN1997D profile of the initially bound 
material for the fiducial model of \citet{Young98} at shock emergence. 
Plots are temperature 
(in units of $10^6\;$\Kelv), density (in units of $10^{-5}\;$\gmcmc) and 
velocity and escape velocity (in units of $10^8\;$\cms). 
\label{fig:initbound}}
\end{inlinefigure}

\subsubsection{Computational Aspects}

Tracking the evolution of the envelope over several years (the expected time 
of emergence of the black hole) is not feasible numerically. Integrating 
the model until emergence would require a CPU time greater than 
one month, even when adjusting the position of the inner boundary and using 
the MTP procedure (see \S~\ref{Sect:code}).
As a result, we devised a rescaling scheme where the original model is 
rescaled to a more rapidly evolving one, while maintaining simple relations 
between the quantities computed for the rescaled model and the original one. 
This rescaling scheme, presented in detail in the appendix, has allowed for 
an additional acceleration of a factor of $\sim5$ in computational 
efficiency, thus reducing the total required computational time to an 
acceptable limit.  

We must note that several particulars of the initial model 
used here for SN1997D imposed some specific complications in employing the 
rescaling scheme. The fact that initially $t_{acc}\la t_0$ 
and that the flow is expected to be moderated at early times 
when a critical accretion rate is approached implies that the 
rescaling scheme described in the appendix cannot be applied at the earliest 
stages of evolution. Furthermore, at early times the energy generation rate 
of the radioactive elements, and especially \Ni, is at its largest. The 
enhancement required for these rates in the course of rescaling (see in the 
appendix) further compounds the applicability of the rescaling scheme at 
very early times. 

Therefore our numerical approach has been to perform the simulation 
of the model in its original scale until three conditions are 
met: 1) the time dependent marginally bound radius is significantly smaller 
(a factor of 5) than the accretion radius; 2) the accretion rate through the 
inner boundary is smaller than 
$0.1 {\dot M}_{crit}$ and 3) that total energy to be emitted in the decays of the 
remaining \Ni is significantly smaller than the internal energy at 
that time. We find that these conditions are all satisfied at the physical 
time of about $t\!=\!25\;$days, which is when the simulation was stopped, 
rescaled by a factor of 5, and restarted. 
We performed the simulation of the rescaled model 
with a single time-step integration until the recombination front has 
settled to a roughly constant Lagrangian position, which occurs after 
about $t\!=\!150\;$days. These first two stages required about 20 CPU hours 
each. From this point on, the simulation is continued with the 
Multi Time-step Procedure, where the integration domain 
was divided into four sub-grids. Performing the simulation up to a physical 
time of $t\!=\!1500\;$days required an additional 120 CPU hours (recall that 
at later times the time step drops as the inner boundary must be moved inward 
in order to maintain it in LTE -- see \S~\ref{Sect:code}). All computations 
were performed on a  SGI 500MHz EV6 machine with a compaq XP1000 processor.

\subsubsection{Numerical Results}

The calculated light curve for SN1997D is shown in 
Fig.~\ref{fig:light97D}. Also shown are the observed data extending up to 
$\sim440$ days after the explosion \citep{Benetti00}, with which the agreement
of the simulated light curve is very good. For comparison, we also show the 
light curve calculated for an identical model but with the abundance of 
radioactive elements reduced by a factor of 300, and also the late-time 
light curve estimated analytically as in Fig.~\ref{fig:analytics} above. In 
the simulation we cannot single-out the accretion luminosity directly, by we 
can evaluate its contribution to the total luminosity by subtracting the 
contributions of radioactive heating. The simulation does record  
the total rate of energy deposition in the envelope by the decays of 
each of the isotopes, $Q_X(t)$, and since we can estimate that the deposited 
heat is emitted as a bolometric luminosity very rapidly after 
thermalization, i.e., $L_X(t)\approx Q_X(t)$,  
we can attempt to identify the accretion luminosity as 
$L_{acc}=L_{tot}-(Q_{^{56}Co}+Q_{^{57}Co}+Q_{^{44}Ti})$.
These calculated contributions for the model are shown 
in Fig.~\ref{fig:tail97D}.

\vspace{0.4cm}
\begin{inlinefigure}
\centerline{\includegraphics[width=1.0\linewidth]{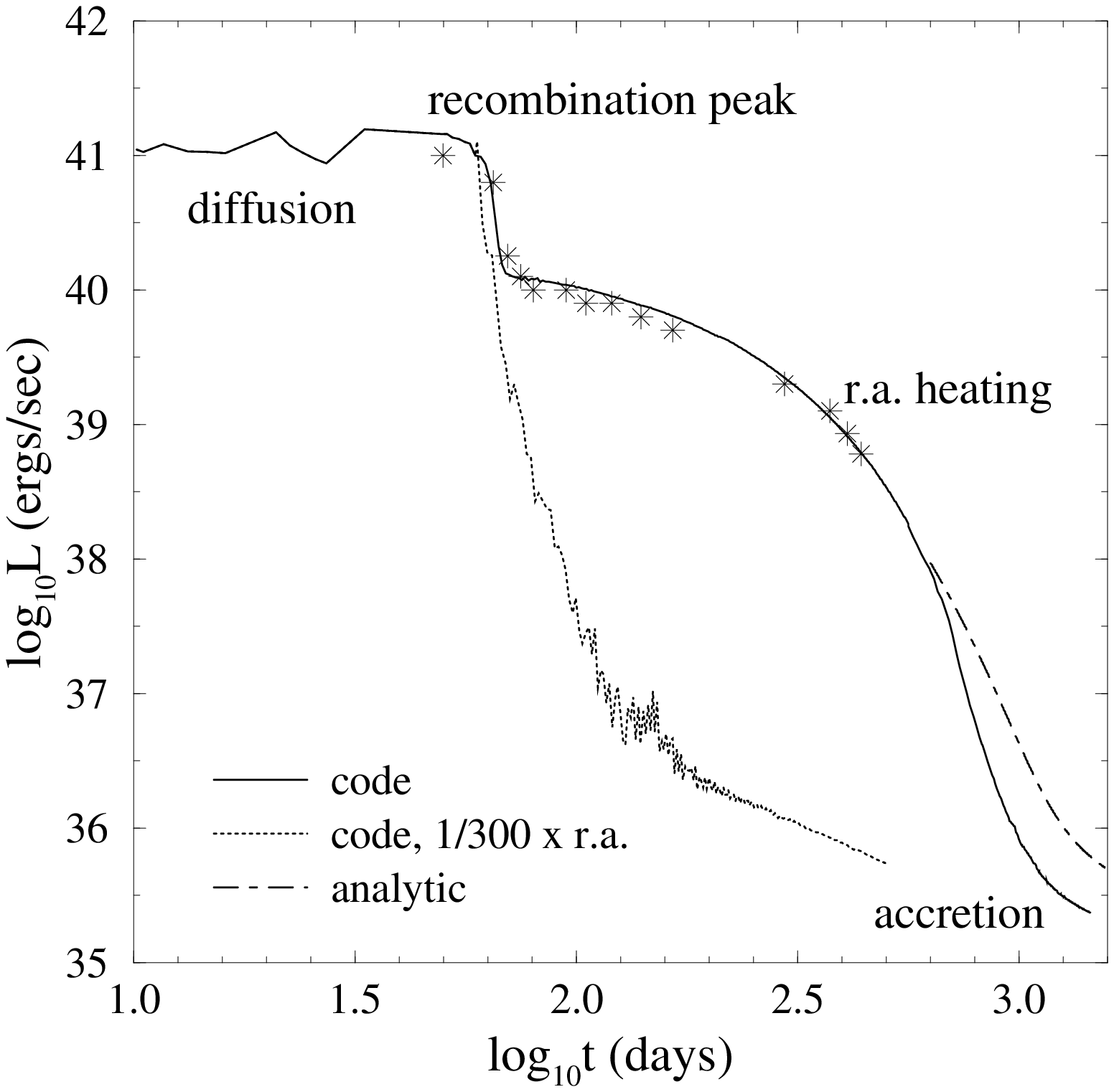}}
\figcaption{\label{fig:light97D}
Numerical result for the bolometric light curve of SN1997D 
using the fiducial initial model \citep{Young98}, and the same model with 
the abundance of radioactive elements reduced by a factor of 300 
(solid and dashed lines respectively). Also shown are the data of observed 
bolometric luminosity for $t\!\leq\!450\;$days (asterisks) from 
\citet{Benetti00}, and the analytic 
estimate of \S~\ref{subsect:97Danalytic} (dot-dashed curve)}
\end{inlinefigure}

The transition from a \Coa-dominated tail in the luminosity is evident 
at about 1000 days. From Fig.~\ref{fig:tail97D} we can determine that 
the accretion luminosity does become the dominant source of the 
light curve, and the time of emergence 
(when $L_{acc}(t_{BH})=\frac{1}{2}L_{tot}(t_{BH}))$ 
is $t_{BH}\approx1050\;$days, 
when $L_{acc}\approx 3.2\!\times\!10^{35}$. 
This is roughly $\frac{1}{5}$ of the accretion luminosity 
(at the same physical time) estimated using the initial 
average values of $\rho_{0,He}=5\!\times\!10^{-6}\;$\gmcmc 
and $t_0=50\;$hrs in equation~(\ref{eq:L_acc}).
This difference is the result of moderation by radiation forces as the 
accretion rate exceeds $\dot{M}_{crit}$, and in part also by the fact that 
initially the envelope's own pressure is not negligible. 
An accretion rate of $\dot{M}>\dot{M}_{crit}$ is actually reached while 
the accretion flow is still building, and the radiation pressure 
it induces causes {\it a significant part} of the 
initially bound material to become unbound. Indeed, the the total accreted 
mass (extrapolated to $t\!\rightarrow\!\infty$) found in the simulation 
is decreased to only $\sim0.067\;M_\sun$, less than a third of the 
originally bound material.

The evolution of the accretion rate is plotted in
Fig.~\ref{fig:mdot97D}, which shows the accretion rate (calculated near 
the inner boundary) as a function of time. The maximum accretion rate is 
reached after about 18 hrs, when its value 
is $\sim 1.5\dot{M}_{crit}$. At this stage the accretion flow is not 
quasi-stationary (see below), and significant moderation is imposed.
We comment that the details of the accretion rate at these very early times 
are dependent on the assumed initial conditions. With initial 
conditions set during collapse, MacFadyen, Woosley, \& Heger (1999) find 
that the earliest accretion reaches a maximum rate as early as a few hundred 
seconds after bounce. In our numerical
simulation the initial conditions are set after the passage of the shock
through the envelope and hence we cannot reproduce these earliest stages of
the accretion history. More significant in the 
context of black hole emergence, however, is the character of the flow as it 
settles on the late-time, dust-accretion track, where there is good agreement 
between our results and those of \citet{MacFWoosHeg99}.

\vspace{0.2cm}
\begin{inlinefigure}
\centerline{\includegraphics[width=1.0\linewidth]{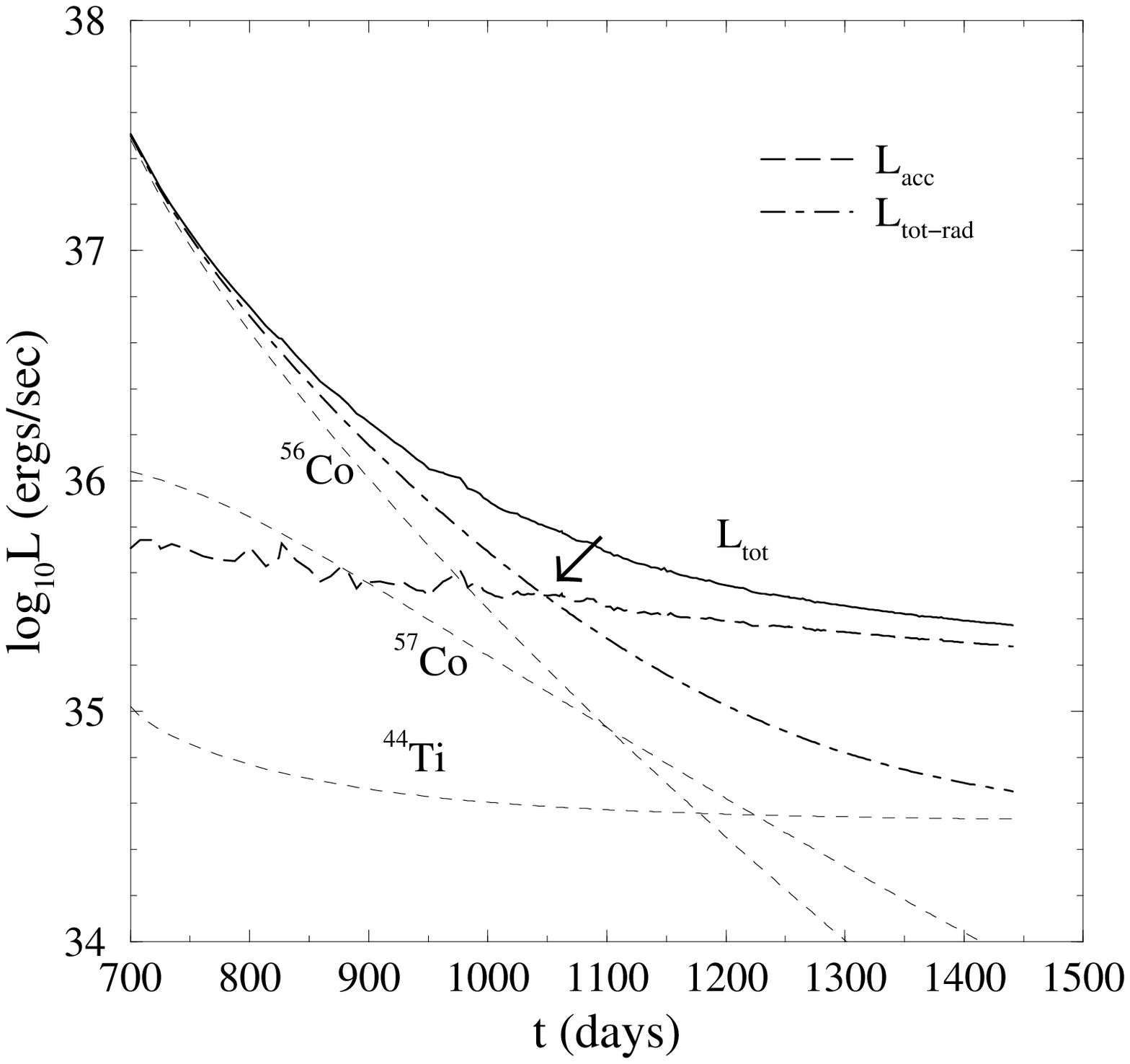}}
\figcaption{The total bolometric luminosity for the nominal model for SN1997D 
and the estimated partial contributions of accretion onto the black hole and 
of radioactive heating during 
over the period $700-1500\;$days from the explosion (Nov.~1998-Dec.~2000). 
The arrow marks the time of emergence of the accretion luminosity 
($L_{acc}=\frac{1}{2}L_{tot}$).
\label{fig:tail97D}}
\end{inlinefigure}

\begin{inlinefigure}
\centerline{\includegraphics[width=1.0\linewidth]{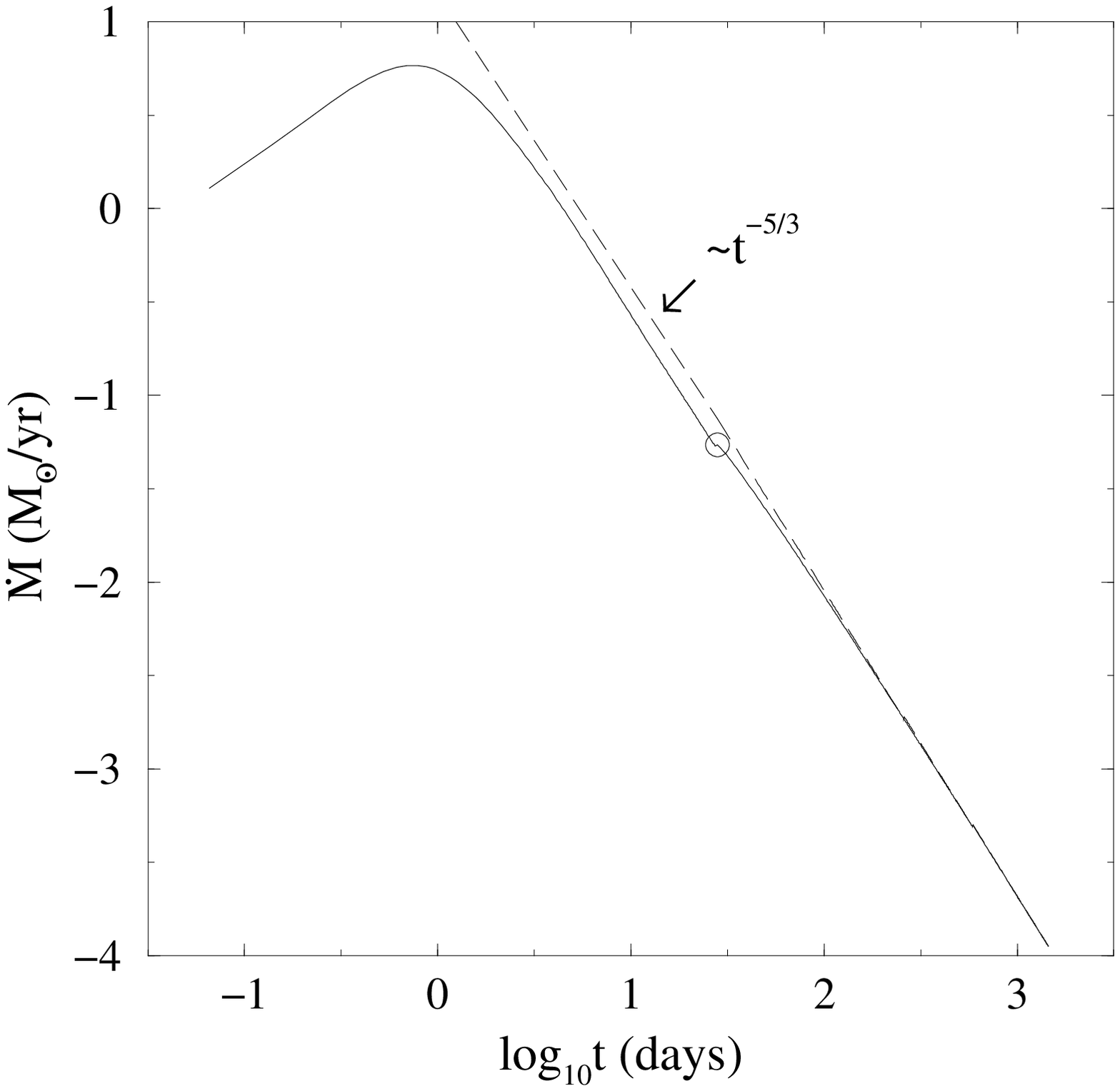}}
\figcaption{Accretion rate ${\dot M}(R_{in})$ close to the inner boundary 
versus time for the fiducial model. The circle denotes the transition from 
the original model to the rescaled one (see text).
\label{fig:mdot97D}}
\end{inlinefigure}
\vspace{0.5cm}

It is apparent that the accretion does eventually settle on a dust-like 
flow $\dot{M}\!\sim\!t^{-5/3}$, as expected, at $\sim10\;$ 
days after the explosion. The quantitative effect of the early history of 
the accretion flow is clearly apparent, since the accretion rate is lower 
than the value predicted by equation~(\ref{eq:dustmdot}) for 
$\dot{M}(10\;\mbox{days})$, based on the initial properties of the inner 
part of the bound region. 
By extrapolating the dust-like accretion rate back in time, we find 
that $\dot{M}=\dot{M}_{crit}$ at $\sim 100\;$hrs, instead of the 
original estimate of $t_{crit}\approx300\;$hrs.

We mark the time of transition between the original model 
and the rescaled one by a circle in Fig.~\ref{fig:mdot97D}. 
After a short transient, the accretion flow resettles on a dust-like 
accretion flow, with a very good fit (accurate to about $10\%$) to the 
initial unscaled model. This agreement suggests that the 
rescaled model is indeed capable of reproducing the accretion flow and 
luminosity.  

In Fig.~\ref{fig:Lvsmdot97D} we show the 
behavior of the computed accretion luminosity as a function of the computed 
accretion rate $\dot{M}$, and
compare it to the analytic estimate of the Blondin formula
(eq.~[\ref{eq:Blondin_L}]). The excellent fit of the simulation to the
analytic ratio demonstrates that the radiation-hydrodynamic evolution
in the accreting region does indeed follow a sequence of quasistationary 
states, so the analytic estimates are also justified in the case of a 
variable chemical composition. The reduced accretion rate is clearly the 
cause for the accretion luminosity being only about one fifth of its value 
(had the entire initially bound mass remained available for accretion).

It is noteworthy that the black hole emerges somewhat earlier than in the 
case of Fig.~\ref{fig:analytics}, in spite of the lower accretion luminosity. 
This is due to a significant complementary effect: the structure of the
helium rich layer reduces the effective optical depth to the high energy 
photons emitted in the radioactive decays. First, the initial column depth 
of the layer is only about $\frac{1}{3}$ of that of a layer with identical 
mass and size but a constant density. Second, the outer part of this layer 
has a larger specific kinetic energy and therefore a smaller $\rho_0 t_0^3$ 
than the average value used in \S~\ref{subsect:97Danalytic}. 
This combination causes the 
$\gamma-$ray column depth of the helium-rich layer to be only 1/5--1/4 of 
that at an identical time for an analytic estimate, and the 
trapping efficiency is decreased accordingly. Correspondingly, the 
contribution of radioactive heating to the bolometric luminosity is 
reduced to what would have been expected if the entire helium-rich 
layer were characterized with $\rho_0=5\!\times\!10^{-6}\;$\gmcmc~ and 
$t_0=50\;$hrs.

The model with reduced radioactive isotopes was investigated in order to 
assess the quantitative effect of radioactive heating on
the dynamics of the flow. This effects can be inferred indirectly from 
Fig.~\ref{fig:light97D}, and directly by comparing the accretion history in 
both models. We find that the accretion histories are practically identical, 
where the late-time accretion rate in the model with 
reduced radioactive heating is larger by only $\sim 5\%$ than in 
the nominal model, implying that 
even in the nominal model the energy deposited by heating hardly affects 
the region of bound material. The calculated accretion luminosities also 
agree to within a few percent, although it is the model with radioactive 
isotopes that suggests a higher luminosity. While this effect may be 
partially a numerical artifact, it may also arise from the slightly larger 
photo-spheric radius found in the original model, presumably caused by 
radioactive heating. We conclude that when the abundance of radioactive 
elements in the ejected envelope is as low as it is in the case of SN1997D, 
it is unlikely to impose a significant effect on the expected accretion rate 
or luminosity. By contrast, even the 
small amount of radioactive isotopes present in the SN1997D envelope is 
sufficient to affect the decline of the recombination peak 
(see Fig.~\ref{fig:light97D}). This effect 
was demonstrated clearly in the investigations of the SN1987A light curve 
\citep{Woosley88,ShigNom90}, and is mainly due to release of heat 
deposited in the inner part of the envelope by \Ni decays. 

\vspace{0.5cm}
\begin{inlinefigure}
\centerline{\includegraphics[width=1.0\linewidth]{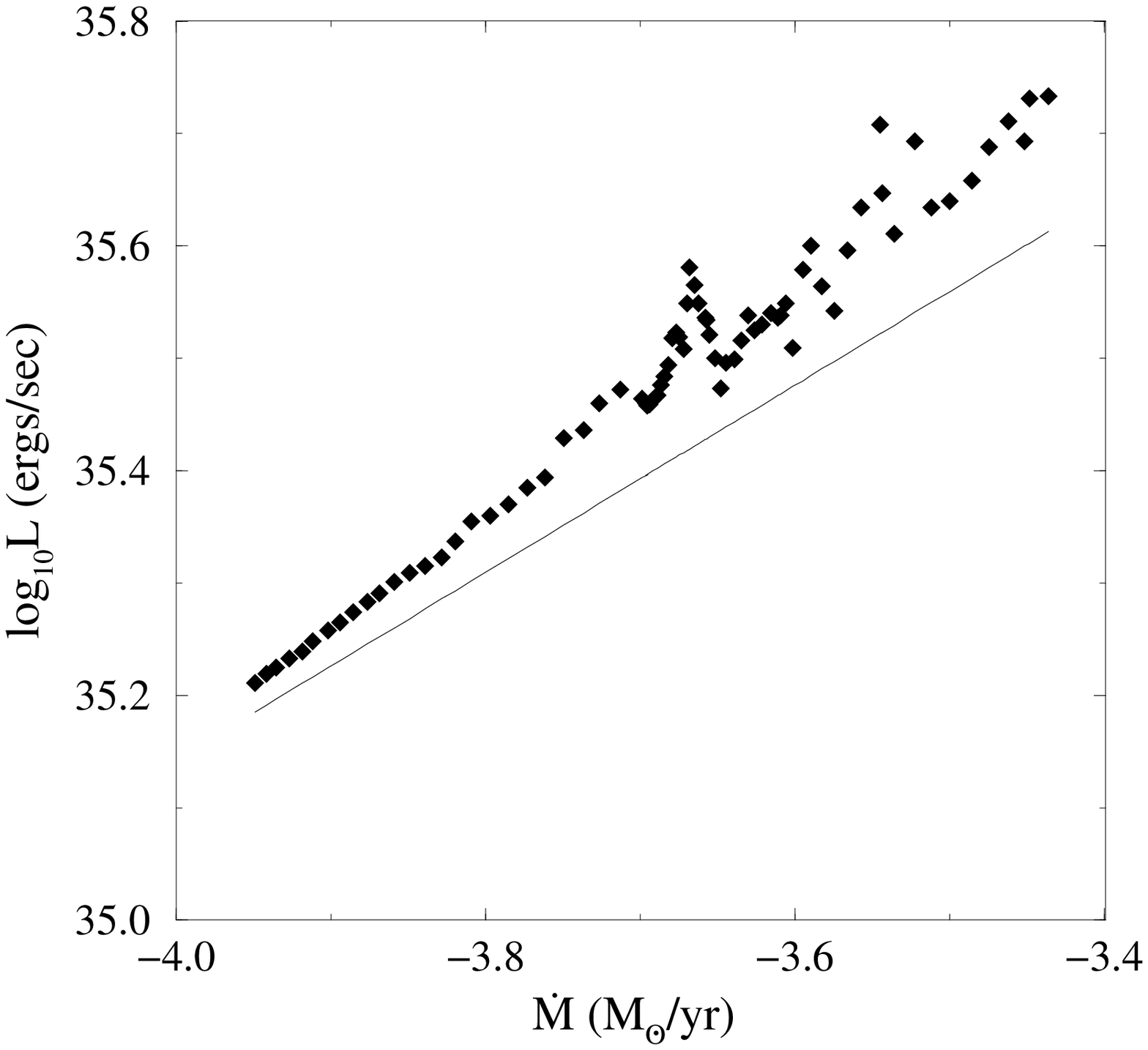}}
\figcaption{The computed accretion 
luminosity (Fig.~\ref{fig:tail97D}) versus $\dot{M}$ (diamonds) for 
the fiducial model of SN1997D. Also shown is the analytic estimate 
based on equation~(\ref{eq:Blondin_L}) for the chemical 
composition of the bound region of the helium-rich layer 
(mass fractions H:0.1, He:0.45, O:0.45). \label{fig:Lvsmdot97D}}
\end{inlinefigure}

The negligible effect of radioactive heating on the hydrodynamic evolution 
allows for a simple extrapolation regarding the expected light curve if the 
abundance of the relevant isotopes is varied. If the abundances of \Cob and 
\Ti are very low, emergence can be gauged by comparing the accretion 
luminosity and the heating by \Coa decays, suggesting that emergence occurs 
at about 970 days. The luminosity at emergence would be slightly higher, with 
$L_{acc}\approx3.9\!\times\!10^{35}$. In this case, the contribution of 
\Coa heating declines rapidly and the late time light curve should indeed 
fall off as a power-law in time. Note that a perfect 
power law decline is found, as expected, for the model with reduced 
abundances of radioactive isotopes.

To conclude our investigation of the SN1997D light curve, we examine the 
history of the luminosity and the accretion flow throughout the envelope.
The profiles of the (comoving frame) luminosity, $L(r)$, and mass flux,
$|\dot{M}|(r)$,
at various selected times during this evolution are plotted in 
Figs.~\ref{fig:Lmdotvsr97D}a and \ref{fig:Lmdotvsr97D}b respectively.    

\vspace{0.5cm}
\begin{inlinefigure}
\centerline{\includegraphics[width=1.0\linewidth]{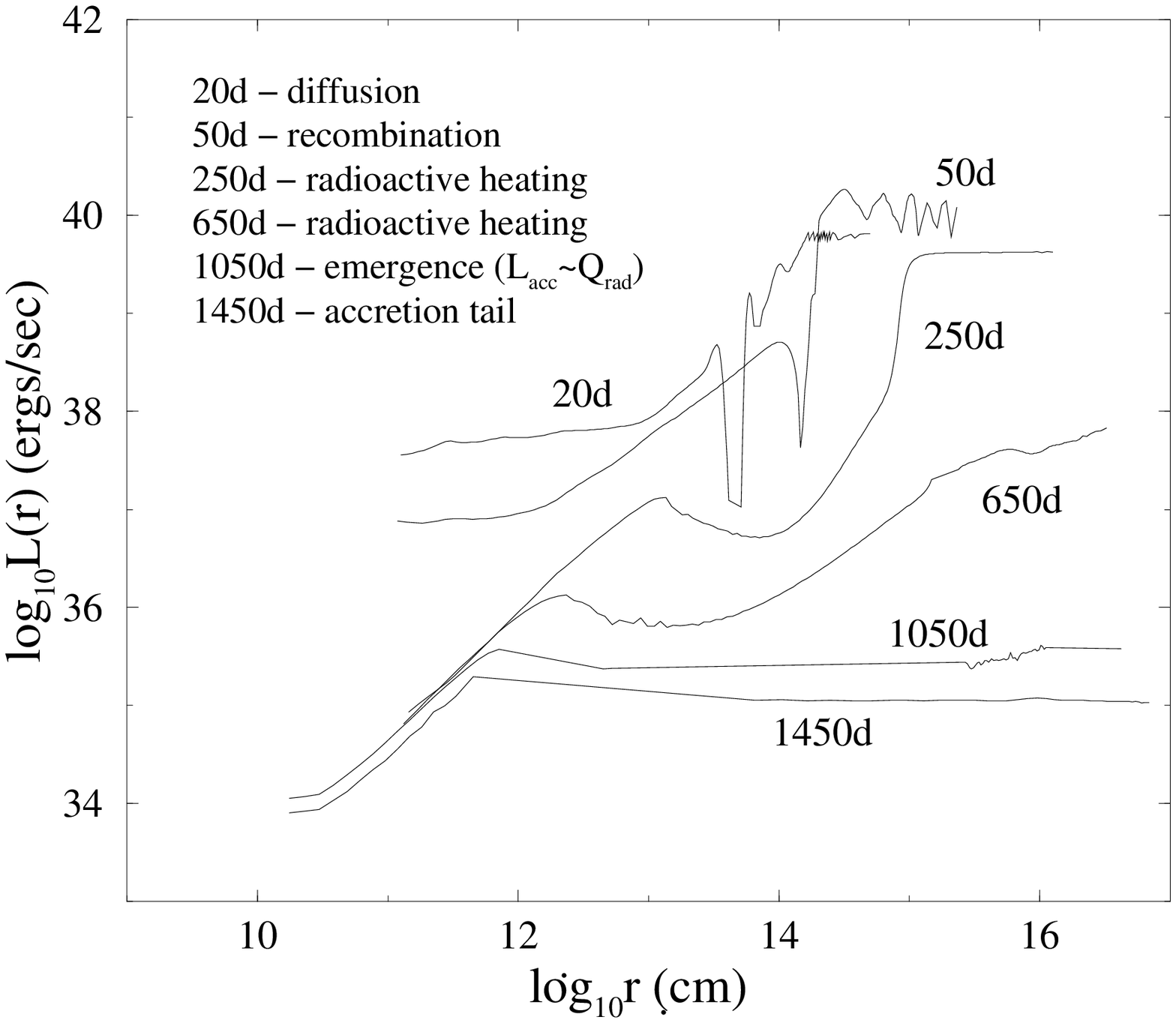}}
\vspace{0.2cm}
\centerline{\includegraphics[width=1.0\linewidth]{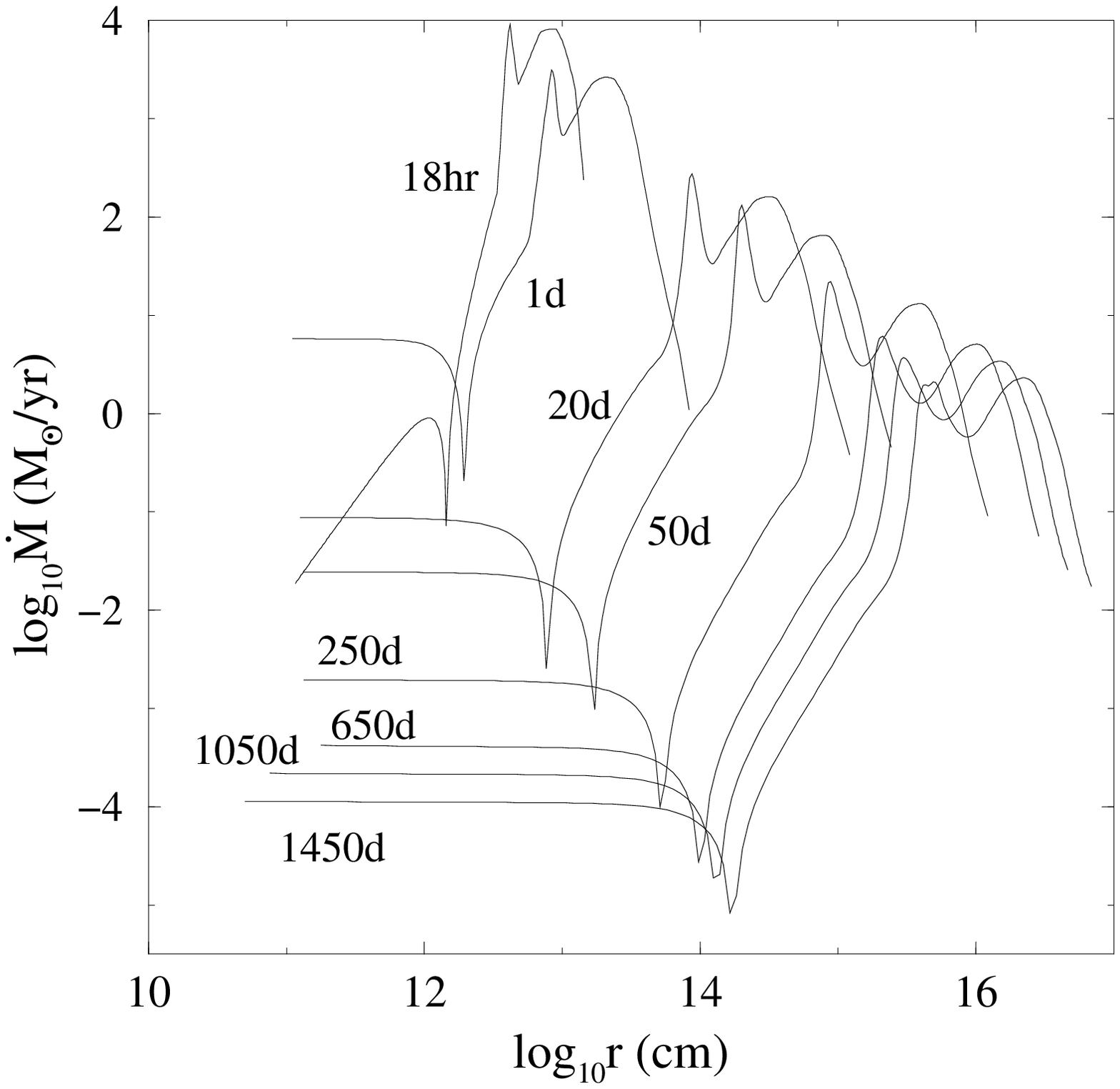}}
\figcaption{Properties of the flow at selected times for the fiducial 
model of \citet{Young98} for SN1997D:
(a) comoving luminosity profiles (b) mass flux profiles.
\label{fig:Lmdotvsr97D}}
\end{inlinefigure}

The luminosity profiles demonstrate the evolution of the light curve through 
its various stages (see Zampieri et al.~1998a for 
details). During the stage dominated by radioactive heating the photo-sphere
lies well above the accreting region. The presence of a separate source 
of luminosity due to compressional heating in the course of accretion  
is evident in the inner part of the flow, but
only when its magnitude becomes comparable to the heating produced by
radioactive decays does it start to dominate the luminosity output.
The drop in luminosity near the inner boundary arises because the
opacity of the partially ionized helium-rich material becomes larger
(much greater than the electron scattering opacity). Closer to the black hole 
the increased temperature increases the degree of ionization of the
helium-rich material, and the comoving luminosity rises again to a 
local maxima. 
Note that the mass flux remains self similar through out the evolution, 
as the accreting region grows in size but the expansion of the envelope 
reduces the magnitude of the accretion rate. 
The sharper, inner spike in the mass flow rate is due to the original spike 
in density of the model at the interface between the helium-rich and 
hydrogen-rich layers (the smoother, outer peak simply corresponds to a 
maximum in the combination $\rho v r^2$).
In the inner region
the accretion rate is nearly independent of radius, which also reflects 
that the accretion flow does indeed follow a sequence of quasi--stationary 
states.

\vspace{0.25cm}
\section{Black Hole Emergence in other Supernovae}\label{Sect:otherSN}
\vspace{0.15cm}

While SN1997D appears to be the only existing candidate for observing 
black hole emergence, theory suggests that the explosions of the most 
massive stars in the core-collapse supernovae range would be significantly 
more favorable for an unequivocal determination of the presence of a 
black hole. The key feature is, of course, the final yield of heavy elements 
in the envelope following early fallback. These heavy elements include 
the radioactive isotopes which mask the accretion luminosity for 
extended times, so that the lower the abundance of radioactive elements, 
the shorter the time until emergence and the larger the luminosity at that time. 
SN1997D may be sufficiently poor in heavy elements so that the accretion 
luminosity can dominate the heating by \Ti decays, but \Coa decays are 
by far the dominant source of power until $\sim 1000\;$days after 
the explosion.        

In a survey of the supernovae of massive stars, \citet{WoosWeav95} concluded 
that explosions of stars of initial masses of $30-40\;M_\sun$ are likely to 
include significant early fallback, thus advecting the newly 
synthesized heavy elements onto the remnant, a likely black hole. For 
``nominal'' explosion energies of $\sim 1.2\!\times\! 10^{51}\;$ergs, they 
found that explosions of stars with masses of $30-40\;M_\sun$ should easily 
lead to remnant masses in the range of $3-10\;M_\sun$, while ejecting 
envelopes practically free of heavy radioactive elements. While 
they did not 
consider explosion energies as low as the $4\times 10^{50}\;$ergs inferred 
for SN1997D, it is natural to assume that lower energy explosions will give 
rise to enhanced earlier fallback and thus higher remnant 
masses\footnote{\citet{WoosWeav95} did examine explosion energies of 
$\gtrsim 2\!\times\!10^{51}\;$ergs. 
In general, they found that if explosion energy is large enough to 
eject a non-negligible amount of radioactive isotopes in stars of 
$30-40\;M_\sun$, the remnant mass will be $\lesssim 2.5\;M_\sun$, and hence, 
possibly, a neutron star}. 

In order to gauge the possible signature of black hole formation in a 
supernova of a very massive progenitor, we computed the 
light curve of such a supernova using the numerical code described in
\S~\ref{Sect:code}. As 
a reference we used a $35\;M_\sun$ progenitor based on model 
S35A of \citet{WoosWeav95}. It has a pre-explosion outer 
radius of $8\!\times\! 10^{13}\;$cm, and is composed of a $14\;M_\sun$ 
helium-rich core and a $21\;M_\sun$ hydrogen-rich envelope. 
The post explosion ejecta profile was constructed so that the total ejected 
masses are 11.5, 12 and 4 $M_\sun$ of hydrogen, helium and oxygen, 
respectively. Most notably, the envelope is assumed to be free of 
radioactive isotopes. The total kinetic energy of the ejecta is set at 
$1.2\!\times\!10^{51}\;$ergs, leaving behind a $7.5\;M_\sun$ black hole. 
Assuming that the helium-rich layer carries 
$1\%$ of the total energy and that the entire envelope has a common 
expansion time scale, our model has $t_0\approx55\;$hrs. The initial 
profile of the bound material thus has 
$\rho_{0,He}=4.93\!\times\! 10^{-6}\;$\gmcmc and therefore  
$\rho_0 t^3_0=3.83\!\times\! 10^{10}\;$\gmcmc$\;\mbox{s}^{-1}$. 
This expansion time scale is significantly shorter than the accretion time 
scale, which for this model is roughly $10\;${\it{days}}, so that the flow 
is dust-like at its onset. The critical time is, however, $\sim280\;$hrs, 
and once again we expect the accretion can begin as dust-like and
become then moderated by radiation pressure when the Eddington limit is
approached. Through the analytic estimates in \S~\ref{subsect:97Danalytic} 
we find that $t_{dust}\approx 110\;$hrs for this model.

The calculated light curve for S35A is shown in 
Fig.~\ref{fig:SN35lite}. We note that no rescaling was required for these 
models due to the relatively large time step and short evolutionary time 
required until emergence; the calculation required
about 55 CPU hours. The modest explosion energy combined with the 
large radius of the progenitor lead to a relatively small initial average 
temperature of about $2.5\!\times\!10^5\;$\Kelv, and recombination occurs 
around a time of $t_{rec}\approx 35\;$days. The light curve during 
this peak is quite similar to the one observed for SN1994W, a reasonable 
candidate for an explosion of a very massive star with very little 
radioactive elements in the ejecta (see below).

\vspace{0.2cm}
\begin{inlinefigure}
\centerline{\includegraphics[width=1.0\linewidth]{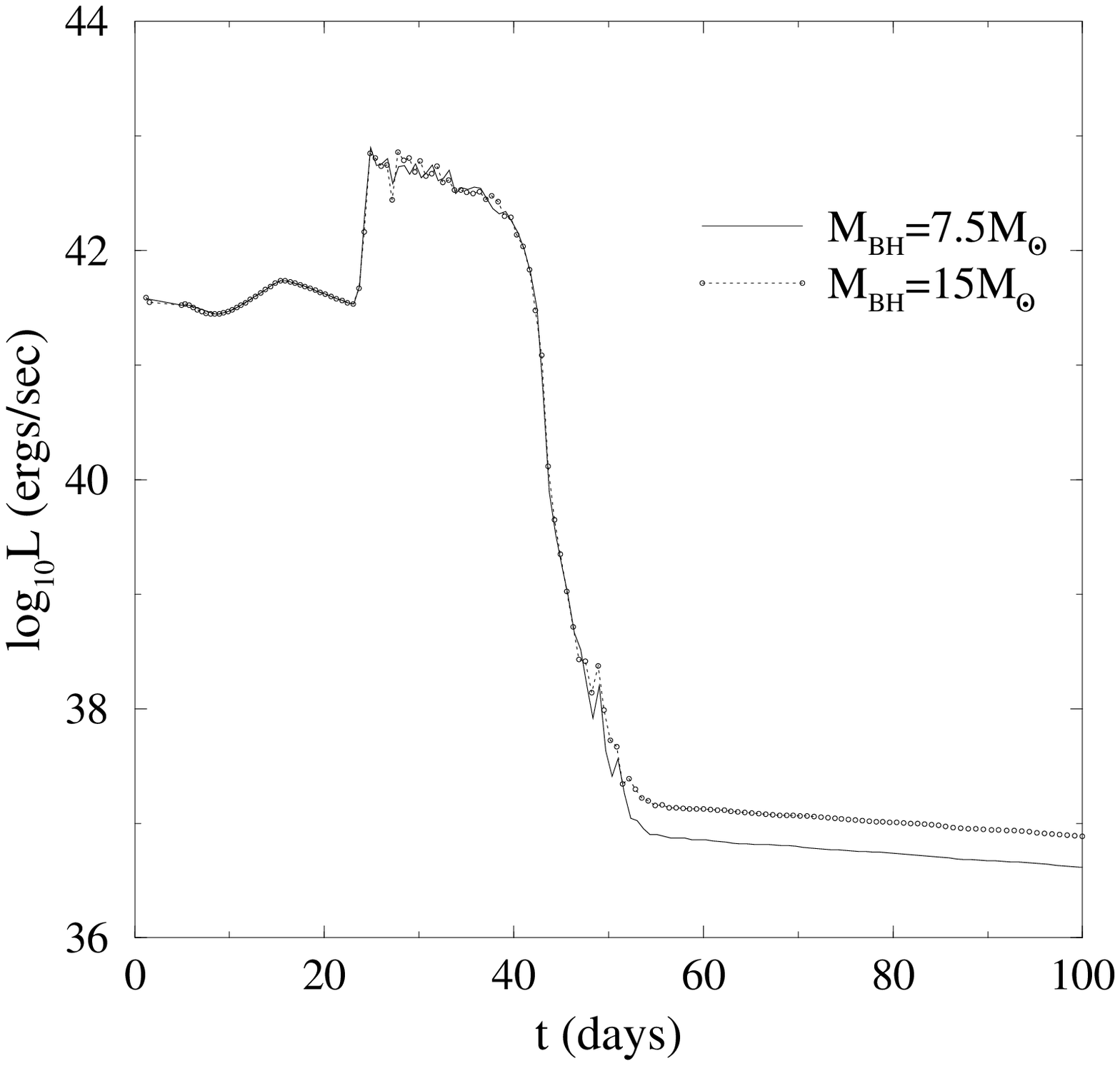}}
\figcaption{The light curve including an accretion tail for the S35A model 
of \citet{WoosWeav95}. Solid line: $M_{BH}=7.5\;M_\odot$, dashed line: 
$M_{BH}=15\;M_\odot$. \label{fig:SN35lite}}
\end{inlinefigure}

In the absence of radioactive heating the recombination peak fades rapidly, 
and the emergence of the black hole occurs at $t\approx 55\;$days with a 
total luminosity at emergence of about $10^{37}\;$\ergss. 
The rapid decline of the recombination luminosity causes the accretion 
luminosity to be nearly the sole contributor to the light curve as 
early as at $t\!\approx\!60\;$days, with a luminosity of 
$L(t=60\;\mbox{days})\approx 7.2\!\times\!10^{36}\;$\ergss. 
This value is {\it smaller} than the analytic result of 
$L(t)=L_{Edd}(60\mbox{days}/t_{dust})^{-25/18}$ by a 
factor of a few, although the calculated accretion rate {\it does} 
satisfy $\dot{M}(t)\simeq \dot{M}_{crit}(t/t_{dust})^{-5/3}$. This 
result is due to the fact that at this stage bound-bound and bound-free 
opacities in the (relatively narrow) partially ionized region in the 
accreting material are high (much larger than free electron scattering 
opacity). This imposes a significant contribution to the total optical depth 
of the accretion flow, increasing its ``effective'' average opacity. 
Note that for a given accretion rate, the accretion luminosity is expected 
to follow a $L_{acc}\propto \kappa^{-1/2}$ dependence 
(using eq.~[\ref{eq:Blondin_L}] and expressing $L_{Edd}$ and $\dot{M}_{Edd}$ 
explicitly). This is a transient effect: as time elapses, expansion of the 
envelope gradually causes the fully ionized, inner part of the accreting 
region to dominate the optical depth of the flow, so that the average 
``effective'' opacity is rapidly declining towards the electron scattering 
value. Consequently, at this stage the luminosity falls off somewhat 
more slowly than a $t^{-25/18}$ power-law, eventually settling on such 
a time dependence.

Also shown in Fig.~\ref{fig:SN35lite} is the same model with the black hole 
mass arbitrarily increased to $M_{BH}=15\;M_\sun$, the light curve 
evolution is identical until emergence, which occurs at a slightly higher 
luminosity and therefore at a slightly earlier time 
($\sim 3\!\times\! 10^{37}\;$\ergss at $t\approx 50\;$days). 
The light curve again evolves thereafter due to accretion
$L(60\mbox{days})\approx 1.33\!\times\!10^{37}\;$\ergss, again falling off 
somewhat more moderately than a $t^{-25/18}$ power law.

We emphasize that the accretion rate and luminosity in both models are 
significantly lower than that predicted by 
equations~(\ref{eq:dustmdot}, \ref{eq:L_acc0}) for the initial parameters 
of the helium-rich layer. Even though the accretion can begin as dust-like, 
the moderation of the accretion flow by radiation pressure when the 
Eddington limit is approached is evident in 
the absolute magnitude of the accretion luminosity-tail.  
The model with the larger black hole mass has a higher critical accretion 
rate, so even though it also induces a smaller $t_{dust}$, the ratio of 
the accretion luminosity is increased by slightly more than just a factor of 
$2^{2/3}$ arising from the ratio of the black hole masses.

We conclude, that {\it Type II supernovae with no radioactive isotopes
provide the most favorable cases for observing the emergence of a black 
hole}. It is noteworthy that a similar explosion of the bare helium core in
Type Ib/Ic supernovae would be far less favorable. Not only will all the 
explosion energy be deposited in the helium layer (instead of only a few 
percent), there will also be no inverse shock that slows
the inner part of the helium layer and enhances fallback (which is essential 
for decreasing the abundance of ejected radioactive elements).  
The very small time step required for simulating such a model prevents 
us from examining it numerically, but we can place some limits 
with the analytic derivations presented in \S~\ref{subsect:97Danalytic}. 
For example, consider a $14\;M_\sun$ helium star with an 
initial radius of $R_0\approx 10^{11}\;$cm exploding with an energy 
$1.2\!\times\! 10^{51}\;$ergs, producing, again, a $7.5\;M_\sun$ black hole. 
We estimate that the post-explosion envelope will have 
$\rho_{0,He} t_0^3\approx 1.85\!\times\! 10^{7}\;$\ergsgm, 
$\rho_0\approx 3\;$\gmcmc and $t_0\approx 0.05\;$hrs.
Even assuming that the flow does begin as dust-like, we find that 
$t_{dust}\approx 4\;$hrs ($t_{crit}\approx 10\;$hrs) so that the luminosity 
at $t\sim60\;$days will be only $\sim 5.3\!\times\! 10^{35}\;$\ergss - 
a few percent of the values found above. While recombination is 
likely to occur earlier in such supernova, the actual accretion rate in such 
an explosion will be strongly modified due to high initial temperatures 
(and pressure), confirming that even radioactive-free  
Type Ib/Ic supernovae are unlikely to be good candidates for observing the 
emergence of black holes. 

\section{Conclusions, Discussion and Observational Prospects}\label{Sect:conc}

In this work we have examined the emergence of luminosity due to 
spherical accretion onto black holes formed in realistic supernovae. 
The study is based on a combination of analytic estimates and 
numerical simulations. Our main focus has been on SN1997D, where the low 
explosion energy and very low inferred abundance of \Coa in its ejecta 
provide a possible (and currently, the only) candidate where the emergence 
of a black hole in a supernova may actually be observable. 

We confirmed the main features of the accretion history and luminosity as 
derived by \citet{PaperI}. At late times, the accretion flow settles on a 
dust-like motion, and the accretion rate falls off as a $t^{-5/3}$ in time 
\citep{CSW96}. Accretion proceeds in
a sequence of quasistationary states, so the accretion luminosity 
achieves the expected magnitude of a hypercritical, spherical 
accretion flow \citep{Blondin86}, accounting for the secular decay in time of 
the accretion rate. The resulting time dependence of the late-time 
accretion luminosity is $L(t)\propto t^{-25/18}$. This relation marks a 
fundamental feature regarding emergence: since the main competitor to 
accretion in powering the late time light curve is radioactive heating which 
decays exponentially with time, emergence of the accretion luminosity is 
inevitable if spherical accretion does indeed persist. While in typical 
Type II supernovae the abundance of radioactive isotopes is expected to be 
large enough to prevent a practical observation of the emergence, 
more rare explosions where this abundance is significantly reduced 
and fallback is enhanced provide important exceptions. 

The case of SN1997D offers a unique opportunity to examine black hole 
emergence \citep{SN97Dlet}, in view of its inferred explosion energy 
of $\sim 4\!\times\! 10^{50}\;$\ergss~ and very low abundance of \Coa, i.e., 
only $2\!\times\!10^{-3}\;M_\sun$ \citep{Turatto98, Benetti00}. We have 
have examined the expected emergence of a black hole in SN1997D analytically 
and numerically. In the analytic study we noted that the early-time 
accretion is almost certain to generate an Eddington-rate luminosity, which 
will modify the accretion history and decrease the late-time accretion rate. 
In the numerical investigation we have included the effects of a variable 
envelope composition. In particular, we incorporated heating due to 
radioactive decays, including a finite (time-dependent) $\gamma-$ray 
optical depth. The wide diversity of time scales involved compelled us to 
use a rescaling scheme in the numerical simulations 
(see in the appendix).

Our various analyses offer a consistent assessment of the emergence of the 
black hole in SN1997D. We confirm the preliminary estimate of 
\citet{SN97Dlet} that a $3\;M_\sun$ should emerge in SN1997D about 
$1000\;$days after the explosion, in contrast to hundreds of years in the 
case of ``standard'' Type II supernovae (like SN1987A) where the abundances 
of radioactive elements are significantly higher. We show that the dominant 
parameter in determining the luminosity at emergence is the amount of kinetic 
energy carried in the post-explosion helium-rich layer (which is the source 
of late-time accretion). For realistic values of this energy, we estimate 
that the total luminosity at emergence will lie in the range 
$5\!\times\! 10^{35}\la L_{tot} \la 3\!\times\! 10^{36}\;$\ergss, emerging at 
$1000-1500\;$days after the explosion. In our numerical study we have 
examined the best fit model of \citet{Turatto98} for SN1997D. The 
simulations suggest that, for this particular model, the black hole emerges 
at about $1050\;$days after the explosion, with 
$L_{tot}\approx 6.5\!\times\!10^{35}\;$\ergss. The results of the simulation, 
which is based on a consistent, fully-relativistic, radiation-hydrodynamics 
code, provides significant support to the analytic approach. Most notably, 
it demonstrates that, after passing through an early 
radiation-pressure-limited stage which reduces the accretion rate, the 
accretion flow does settle on the dust-like solution and that the influence 
of the radioactive decays on the hydrodynamic 
evolution is negligible for the low abundances inferred in SN1997D.  

The time of emergence and the corresponding luminosity depend on the 
abundances of \Cob and \Ti in the envelope, which are unknown a-priori. 
However, as long as the accretion luminosity is significantly larger than 
the threshold set by positron emission in \Ti decays, both emergence time 
and luminosity will not vary substantially. At the time of 
emergence $\gamma-$ray escape from the envelope is non-negligible and 
increases rather rapidly with time, as the envelope's optical depth drops. 
Correspondingly, the bolometric luminosity due to radioactive heating 
decreases even faster than exponentially with time. 

Although our results appear to be quite robust within the context of the 
present study, some uncertainties arise due to effects not considered here. 
Most critical is the assumption of perfectly spherical accretion: if the 
bound material can maintain sufficient angular momentum, a centrifugal 
barrier would drive the flow into a disk structure \citep{Chevalier96}. 
The radiative efficiency of 
hypercritical disk accretion is uncertain, complicating any a-priori estimate 
regarding the character of the late-time accretion luminosity. 
The residual angular momentum is likely to be dependent on the 
details of the progenitor and the explosion 
(we note that for SN1987A the observed neutrino burst seemed 
consistent with nonrotating models \citep{Burrows88}, 
indicating that angular momentum of matter close to the collapsed core did 
not play a significant role). In principle, late time accretion occurs many 
viscous time scales after the explosion, and much of the initial angular 
momentum could be lost, but the minimum specific angular 
momentum required to force disk accretion is quite small (about 
$(GM_{BH} R_{ISCO}c)^{1/2}$, where $R_{ISCO}=6GM/c^2$ is the innermost-stable 
circular orbit near a black hole). 
Furthermore, even if at late times angular momentum is negligible, some 
residual effect may arise if the character of the early-time accretion has 
been sufficiently altered (Mineshige et al. 1997 found through simulations 
that sufficient angular momentum strongly affects the magnitude and time 
dependence of the accretion during the first {\it hours} after collapse). 
We defer further investigation of fallback including angular momentum to 
future work. 

Additional effects could arise due to convection, mixing and clumping of 
envelope material (and radioactive isotopes in particular), and dust 
extinction, although 
we expect that at a time of $\gtrsim3\;$years after the explosion, any 
dust formed will have become transparent to the underlying luminosity.   
This may not be true if photoionization by the central accreting black hole
(not included in our calculation) is important.

Our results illuminate the uniqueness of SN1997D.  
It is the low abundance of radioactive elements, combined with the 
low explosion energy in general, that allows the black hole to emerge only 
a few years after the explosion, at which time the absolute magnitude of the 
accretion luminosity is still quite large (recall that for ``standard'' 
explosions we expect emergence luminosities of 
$10^{32}-10^{33}\;$\ergss). Since gas
and radiation are in LTE in the inner accreting region and the photosphere
is well localized in radius, the emerging spectrum is expected to be very
similar to a blackbody at the photospheric temperature ($\sim 7500$ K)
with the superposition of heavy line elements. We therefore expect that 
emission will be mainly in the optical band.
At a distance of $\sim 14\;$Mpc, the 
corresponding apparent magnitude for the range of emergence luminosities 
predicted for SN1997D is $m_V\approx 28-30$ and $m_R\sim m_I \approx 27-29$.
The higher luminosity (lower magnitude) end of this range 
coincides with the detection threshold of the HST STIS camera, indeed 
making emergence in SN1997D potentially {\it observable}. We note that 
the capabilities of HST are needed both because of the faintness of the 
source at this time, and the resolution, required to distinguish the 
supernova over the background of the parent galaxy NGC 1536. A recent 
estimate (M.~Turatto 1999, private communication) suggests that the required 
exposure time for the HST STIS to resolve the apparent magnitude expected at 
emergence in SN1997D at a signal to noise (S/N) ratio of 10 would 
be $\sim 25000\;$seconds. The accretion luminosity should be 
distinguishable from radioactive heating (or from circumstellar interaction, 
which has not been observed in SN1997D so far, Benetti et al.~2000) due to 
its unique power law dependence. This would be especially true if the 
abundances of \Cob and \Ti are negligible, since after emergence 
the luminosity would be due only to accretion. But, even if there is a finite 
contribution from these two isotopes to the light curve, the presence of a 
power-law source may still be identifiable if several measurements with 
reasonable accuracy could be made. The novel consequences of a positive 
observation are self-evident - the first {\it direct} observational evidence 
that a black hole can be formed by an otherwise ``successful'' supernova.               
Although our main focus has been SN1997D, other theoretical Type II 
supernova models offer more favorable conditions for an actual 
observation of black hole emergence. 
The survey of \citet{WoosWeav95} suggests that progenitors with masses in 
range $30-40\;M_\sun$ are quite likely to leave behind a remnant 
of mass $3-10\;M_\sun$ and a practically radioactive-free envelope, as 
all the iron group isotopes are engulfed by the early fallback. Our 
calculations in \S~\ref{Sect:otherSN} show that accretion 
luminosity will emerge after recombination depletes the initial internal 
energy of the envelope, hence after {\it a few tens of days}. The 
corresponding luminosity at emergence will be $\sim 10^{37}\;$\ergss, 
which with the present capabilities of HST would allow for detection 
at emergence perhaps out to $20-25\;$Mpc. The luminosity at 
emergence is higher than for SN1997D so that the galactic background would be 
less important, and ground based instruments might be relevant as well.

How often can we expect an opportunity to observe a black hole emerging in a 
supernova? An upper limit is, of course, the rate at 
which black hole forming supernovae occur. 
Adopting the recent results of \citet{FryKal99} 
that about $10\%$ of core collapse supernovae make black holes, we 
expect a rate of about one event per thousand years per galaxy. 
There are somewhat less than one thousand galaxies at a distance 
$\la 20\;$Mpc from the Milky way, and hence there may be approximately one 
observable event (equal or better than SN1997D) per several years. This 
estimate is an upper limit, 
since a significant fraction of high mass progenitors will end their lives 
as Type Ib/Ic supernovae after losing their hydrogen envelope, and such 
supernovae are unfavorable for a practical observation at emergence. 
Furthermore, even ``perfect'' candidates may not allow for an actual 
observation, as seems to have been the case in SN1994W. 
Following its recombination peak, the tail of the SN1994W light curve failed 
to settle on an exponential decay that would arise due to \Coa decay 
\citep{Sollerman94W}. Interestingly, the last data point led to a 
limit of $\le2\!\times\!10^{-3}$ on the mass of \Coa in the ejected envelope. 
However, the light curve showed a very steep power-law in time until 
it dropped below the detection threshold some 220 days after the explosion. 
The luminosity during this phase was much too high to be due to accretion 
($10^{40}-10^{42}\;$\ergss) and \citet{Sollerman94W} found that it was most 
likely caused by interaction with circumstellar material, with dust 
formation also playing an important role. We can only 
speculate that in the absence of circumstellar interaction and dust 
extinction, the light curve would have declined exponentially after the 
recombination peak and the accretion luminosity may have emerged while still 
marginally detectable. The case of SN1994W implies that the rate of potential 
observable black hole emergence may be significantly smaller than 
that of black hole forming Type II supernovae. 

Nonetheless, we urge that any Type II supernovae which shows a diminished 
\Coa abundance should be monitored, and its light curve be tracked
for several months 
after the explosion. The prospects of observing black hole emergence may 
improve significantly with potential future instruments, such as NGST or a 
dedicated faint-supernovae project.

\acknowledgements

We are grateful to Tim Young for providing us with the results of 
early-time simulations for SN1997D and to Chris Fryer, Robert Kirshner, 
Phillip Pinto and Massimo Turatto for useful discussions. We thank
Monica Colpi and Roberto Turolla for carefully reading the manuscript. 
This work was supported by NSF Grants 
AST 96-18524 and PHY 99-02833, and NASA Grants NAG 5-7152 and NAG 5-8418 
at the University of Illinois at Urbana-Champaign. L.~Z.~acknowledges 
finacial support from the Italian Ministry for University
and Scientific Research (MURST) under grant 
cofin--98--2.11.03.07 at the University of Padova.


\appendix
\section{The Rescaling Scheme: Radiation-Hydrodynamic Renormalization}
\label{Sect:rescale}

The inclusion of the accreting region imposes an inherent obstacle in any 
numerical simulation of a supernova light curve. In the innermost accreting 
region the material is free-falling at a velocity which is a significant 
fraction of the speed of light, so that the dynamical time scale is less than 
$10^{-3}\;$sec. On the other hand, the evolutionary time for the light curve 
is of the order of tens of days to clear the recombination phase, and years 
to reach black hole emergence if there is a non-negligible amount of 
radioactive elements in the envelope. This extreme diversity of time scales 
prevents any single calculation which tracks the entire envelope.

As described by \citet{PaperI}, a considerable increase in the shortest 
physical time scale (and, hence, the Courant integration time step) can be 
obtained moving the inner boundary, $R_{in}$, outward (at hundreds to 
thousands of Schwarzschild radii, but always {\it well within} the accreting 
region). This turns out to be possible because the gas in the inner accreting
region is in LTE and in free-fall, thus allowing a reliable setting of
the boundary conditions even if $R_{in}$ is much larger than the black hole
horizon. When a Lagrangian zone passes through the inner boundary, it is
removed from the calculation and the remainder of the envelope quantities
are rezoned to maintain a satisfactory resolution.

Even with such a relocation of the inner boundary, the feasibility of a 
numerical computation requires some additional measures to accelerate the 
computation. To this end, the code incorporates the Multi-Time-Step Procedure,
where the integration domain is separated into several subgrids, and each
is evolved at its own local time scale; most of the CPU time is thus devoted
to evolving the regions with smaller time steps. We note that this
procedure requires a communication scheme between the sub-grids based on
extrapolation, and can be safely used when there are no sharp
features propagating through the envelope. Thus, we use it only after that 
the recombination front has completed its sweep of the envelope material. 
The speed-up allowed by the MTP is typically a factor of 5-8.

Unfortunately, moving the inner boundary outward and using the MTP is not 
sufficient when the target physical time is of the order of 
years\footnote{We note that using a fully implicit code would not allow a 
significant gain in computational time because limitations on the time-step 
imposed by accuracy are comparable to limitations imposed by the Courant 
condition throughout most of the evolution.}. In the case of SN1997D, 
tracking the evolution of the envelope for $\sim1000\;$days would require 
roughly 1000 CPU hours on the SGI 500MHz EV6 compaq XP1000 processor which 
we have used for the simulations. We were hence compelled to develop an 
additional acceleration algorithm, which is a rescaling 
scheme that allows us to integrate to the black hole emergence stage within a 
reasonable computational time.                          

\subsection{Principles of Rescaling}

The main objective in devising a rescaled model of a specific supernova 
evolution is to maintain all the quantitative features of the realistic 
model, while reducing the required computational time. 
The code uses an explicit Lagrangian scheme. Since the photon transport 
is treated by actually solving the transport equation, the reference 
velocity is the speed of light, and the Courant time step is determined by 
the width of the thinnest zone. This is always the innermost zone,  
$\Delta R_{in}(t)$, which at any time $t$ is a given fraction of the 
position of the inner boundary, $R_{in}(t)$. The efficiency of the code is 
essentially determined by comparing the (small) ratio of the time step to a 
characteristic evolutionary time -- 
for example, the expansion time scale, $t_0$ -- 
\begin{equation}\label{eq:effcode}
\frac{\Delta R_{in}(t)/c}{t_0}\propto\frac{R_{in}(t)/c}{t_0}\;. 
\end{equation}
The essence of a rescaling scheme would then be devising a rescaled model 
where \begin{equation} \label{eq:rescaler}
\frac{R_{in}^s(t^s)/c}{t_0^s}=\alpha\frac{R_{in}(t)/c}{t_0}\;, 
\end{equation}
where a subscript $s$ denotes the rescaled quantities. The rescaled model 
is then more efficient by a factor of $\alpha>1$. 

Successful rescaling must (a) provide a simple relation between its own 
physical quantities and that of the realistic model and (b) maintain the 
hierarchy (inequalities) of time and energy scales that exists in the 
realistic model, so that the relative importance of the different physical 
process is carried on into the rescaled model. The rescaling scheme we have 
chosen to employ in our simulations is detailed below.

\subsection{The Rescaling Relations}

A natural rescaling scheme would be one where all the physical quantities 
of a mass element ($r,v,T,\rho...$) of the realistic 
model are mapped with simple linear relations to the 
rescaled model, i.e., $r^s=\alpha_R r,\;v^s=\alpha_V v$, etc.
In our rescaling scheme we distinguish between the local thermodynamic and 
kinematic quantities of the mass elements in the envelope. Since the photon 
opacities are sensitive and complex functions of the thermodynamic quantities 
, i.e., density and temperature, we find it preferable to leave these 
quantities unchanged in the course of rescaling $T^s\!=\!T,\;
\rho^s\!=\!\rho$, which leads to $c_s^s\!=\!c_s,\;\kappa^s\!=\!\kappa$. 
Note that such a choice also implies that the inner boundary of the 
realistic model should approximately map onto the inner boundary in the 
rescaled model, since the position of the inner boundary is determined by the 
requirement of LTE, which arises from the local values of temperature and 
density.

Once this choice has been made, the constraints on the rescaling scheme are 
quite straight forward. If we require that the ratio of diffusion and 
recombination times scales to the expansion time scale remain unchanged, i.e,
\begin{equation}\label{eq:restimes}
\frac{t_{diff}^s}{t_0^s}=\frac{t_{diff}}{t_0}\;,
\frac{t_{rec}^s}{t_0^s}=\frac{t_{rec}}{t_0}\;,
\end{equation}
the rescaled model must satisfy (see eqs.~[\ref{eq:t_diff},\ref{eq:t_rec}]):
\begin{equation} \label{eq:resRV}
R_0^s V_0^s=R_0 V_0\;.
\end{equation}

The condition (\ref{eq:resRV}) is sufficient to determine the rescaling 
strategy when consistently mapping all radii between the realistic and the 
rescaled models. If all radii are decreased by a factor of $\alpha$, 
all velocities must be increased by the same factor; the efficiency of the 
simulation (eq.~[\ref{eq:effcode})]) is then increased by a factor of 
$\alpha^{-1}/\alpha^{-2}=\alpha$.

If we further impose invariant luminosity ratios according to
\begin{equation}\label{eq:reslums}
\frac{L_{diff,0}^s}{L_{acc,0}^s}=\frac{L_{diff,0}}{L_{acc,0}}\;,
\frac{L_{rec}^s}{L_{acc,0}^s}=\frac{L_{rec}}{L_{acc,0}}\;,
\end{equation}
we arrive (using eqs.~[\ref{eq:L_diff0},\ref{eq:L_rec} and \ref{eq:L_acc0}]) 
at the condition
\begin{equation}\label{eq:resRM}
M_{BH}^s R_0^s=M_{BH} R_0\;. 
\end{equation}
The principle relations of our rescaling scheme are thus
\begin{equation}\label{eq:resratios}
\left(\frac{R_0^s}{R_0}\right)^{-1}=\left(\frac{V_0^s}{V_0}\right)=
\left(\frac{M_{BH}^s}{M_{BH}}\right)=\alpha\;.
\end{equation}
This choice of rescaling allows us to consistently map the marginally bound 
shell of the envelope, since the marginally bound radius, $R_{mb}$ also 
rescales as $R_{mb,0}^s\!=\!(2 M_{BH}^s (t_0^s)^2)^{1/3}=\alpha^{-1}R_{mb,0}$.

Further relations that immediately follow are 
\begin{equation}\label{eq:restLMMdot}
\left(\frac{t^s}{t}\right)=\alpha^{-2}\;,\;
\left(\frac{L_{acc,0}^s}{L_{acc,0}}\right)=\alpha^{-1}\;,\;
\left(\frac{M^s_{end}}{M_{env}}\right)=\alpha^{-3}\;,\;
\left(\frac{\dot{M}^s}{\dot{M}}\right)=\alpha^{-1}\;.
\end{equation}

The efficiency of rescaling is explicit in the rescaling of time -- in the 
rescaled model all physical times are reduced by a factor of 
$\alpha^2$, while the {\it numerical} time step is only reduced by a factor 
of $R_{in}^s/R_{in}\!=\!\alpha$.
 
\subsection{Limitations of Rescaling}

The rescaling scheme presented above is not without limitations; most 
notably, the acceleration factor, $\alpha$ is limited to a value of a few, 
as discussed below. 
Nonetheless, even a value of $\alpha\!=\!5$, which was used in the 
case of SN1997D reduces the required computational time from weeks 
to days, hence allowing for a practical numerical investigation.

\subsubsection{The Acceleration Factor}

Since the rescaling scheme calls for an increase of a factor of $\alpha$ of 
the envelope velocities, clearly $\alpha$ cannot be arbitrarily large, 
since we must maintain a sub-relativistic envelope. 
Furthermore, since the black hole mass is increased, its 
Schwarzschild radius increases 
as well, and naturally it should not be allowed to approach the integration 
domain throughout the calculation. Both these constraints result in the 
practical limit on the value of $\alpha$ to no more than a few.

\subsubsection{Radioactive Elements}

The scaling laws of luminosity and mass in equation~(\ref{eq:restLMMdot}) 
create an undesired inconsistency regarding the luminosity from radioactive 
heating, since for this source $L\propto M$. In order to recover the correct 
rescaled power output from radioactive decays, the energy generation rate 
from radioactive decays per unit mass in the envelope 
must therefore be increase by a factor 
of $\alpha^2$. This could potentially impose a significant deviation in the 
evolution of the rescaled model from that of the original one, if it 
artificially enhances the importance of energy deposition from radioactive 
decays when compared to the other energy scales. However, since we limit 
ourselves to envelopes where the abundance of radioactive elements is 
very small, the artificial amplification of the energy production per unit 
mass of radioactive decays has a negligible effect.

It is further noteworthy that the effective transparency of the photons to 
the $\gamma$-ray photons emitted in the decays must be adjusted in order 
to recover the appropriate scaling of the $\gamma-$transparency time. 
The simple form of $\gamma$-ray opacity (eq.~[\ref{eq:t_trnsgam}]) 
allows to achieve consistency with the change 
$\kappa_\gamma^s=\alpha \kappa_\gamma$, so that 
$t_{trans,\gamma}^s/t_0^s=t_{trans,\gamma}/t_0$. 
A similar transformation for the optical 
photons would be more complex, but can be neglected since the onset of 
thermal-photon transparency is imposed by recombination (rather 
than expansion), and is chiefly determined by the thermodynamic variables 
$\rho$ and $T$, which are kept invariant.
      
\subsubsection{The Accretion Time Scale and Radius}

A second inconsistency that arises in this particular rescaling scheme 
concerns the initial accretion time and radius, $t_{acc,0},\;R_{acc,0}$. 
As is evident from equations~(\ref{eq:t_acc0}, \ref{eq:R_mb-R_acc}), when the 
thermodynamic properties of the helium layer are not changed, so that the 
sound speed is conserved, $t_{acc},R_{acc}\propto M_{BH}$. Our rescaling 
scheme actually {\it increases} these quantities by a factor of $\alpha$, 
while other characteristic times and radii are {\it decreased} by factors of 
$\alpha^2$ and $\alpha$, respectively. This is an inevitable consequence of 
our choice of rescaling.

This inconsistency is significant, however, only while the accretion flow is 
Bondi (1952)-like, when $t_{acc}\leq t_0$ and $R_{acc}\leq R_{mb}$ 
\citep{CSW96}. As the envelope continues to expand, the hierarchy of 
time scales and radii is eventually reversed and the accretion 
eventually transforms into a dust-like flow. Since the accretion time does 
not  enter explicitly in any of the estimates for evolution and luminosity 
(see \S~\ref{Sect:basics}), it is only this hierarchy which is of real 
consequence, and hence rescaling can be applied safely to models which 
are characterized by $t_{acc}\gg t_0$, $R_{acc}\gg R_{mb}$.
     
\subsubsection{Kinetic vs. Thermal Energies}

A similar analysis concerns the ratio of kinetic and thermal energies in the 
envelope. It is clear that the kinetic energy will scale as 
$E_{kin}\!\sim\! M V^2 \rightarrow E_{kin}^s\!/\!E_{kin}\!=\!\alpha^{-1}$, 
while the internal energy (neglecting radioactive heating) will scale as 
$E_{th}\!\sim\! R^3 \rightarrow E_{th}^s\!/\!E_{th}\!=\!\alpha^{-3}$, since 
the temperatures are unchanged. Rescaling thus amplifies the relative 
importance of kinetic energy over thermal energy in the expanding envelope.

However, as is the case for the accretion and expansion time scales, the 
real issue is the hierarchy of energy scales. In fact, the assumption of 
homologous expansion is equivalent to assumption that the conversion of 
thermal energy to kinetic energy is negligible, i.e., 
that $E_{kin}\gg E_{th}$, which is satisfied in the context of our rescaled 
models.  
    
\subsection{A Test of Rescaling}

\begin{deluxetable}{c c c c c c c c c c c c c c}
\tablefontsize{\scriptsize}
\tablewidth{0pt}
\tablecaption{Parameters for the toy model used to test rescaling 
\tablenotemark{\star}
\label{tab:toymodel}}
\tablehead{
\colhead{model} &
\colhead{$M_{BH}$} &
\colhead{$M_{tot}$} & 
\colhead{$M_{He}$\tablenotemark{a}} & 
\colhead{$M_{H}$\tablenotemark{b}} &
\colhead{$R_{out}$\tablenotemark{c}} &
\colhead{$R_{He}$\tablenotemark{d}} &
\colhead{$V_{out}$\tablenotemark{e}} &
\colhead{$M_{ra}$\tablenotemark{f}} &
\colhead{$\varepsilon_{ra}$\tablenotemark{g}} &
\colhead{$\tau_{ra}$\tablenotemark{h}} &
\colhead{$\kappa_{ra}/Y_e^T\;\;$\tablenotemark{i}} &
\colhead{$t_0$} &
\colhead{$t_{acc,0}$}\\
\colhead{} &
\colhead{$(M_\sun)$} &
\colhead{$(M_\sun)$} &
\colhead{$(M_\sun)$} &
\colhead{$(M_\sun)$} &
\colhead{(cm)} &
\colhead{(cm)} &
\colhead{(\cms)} &
\colhead{($M_\sun$)} &
\colhead{(ergs gm$^{-1}$ s$^{-1})$} &
\colhead{(days)} &
\colhead{(cm$^2\;$gm$^{-1}$)} &
\colhead{(hrs)} &
\colhead{(hrs)} 
}
\startdata
Original & 3     & 18.75 & 3.75 & 15  
         & $2\times 10^{13}$ & $4.5\times 10^{12}$ & $6.67\times 10^8$ 
         & 0.036 & $9\times 10^{8}$ & 50 & 0.01 
         & 25/3 & 50 \\
Rescaled & 15   & 0.15  & 0.03 & 0.12 
         & $4\times 10^{12}$ & $9\times 10^{11}$ & $3.35\times 10^9$ 
         & $2.88\times 10^{-4}$ & $2.25\times 10^{10}$ & 2 & 0.05  
         & 1/3 & 250 \\
\tablenotetext{\star}{Rescaled by a factor $\alpha=5$ (see text)}
\tablenotetext{a}{mass of helium-rich layer}
\tablenotetext{b}{mass of hydrogen rich layer}
\tablenotetext{c}{outer radius}
\tablenotetext{d}{outer radius of helium-rich layer}
\tablenotetext{e}{initial velocity at outer radius}
\tablenotetext{f}{total mass of radioactive isotope in the envelope}
\tablenotetext{g}{energy generation rate per unit mass of radioactive isotope}
\tablenotetext{h}{life-time of radioactive isotope}
\tablenotetext{i}{$\gamma-$ray opacity per electron radioactive isotope}
\enddata
\end{deluxetable}

\begin{figure*}[htb]
\begin{minipage}[t]{70mm}
\begin{center}
\includegraphics[width=8cm]{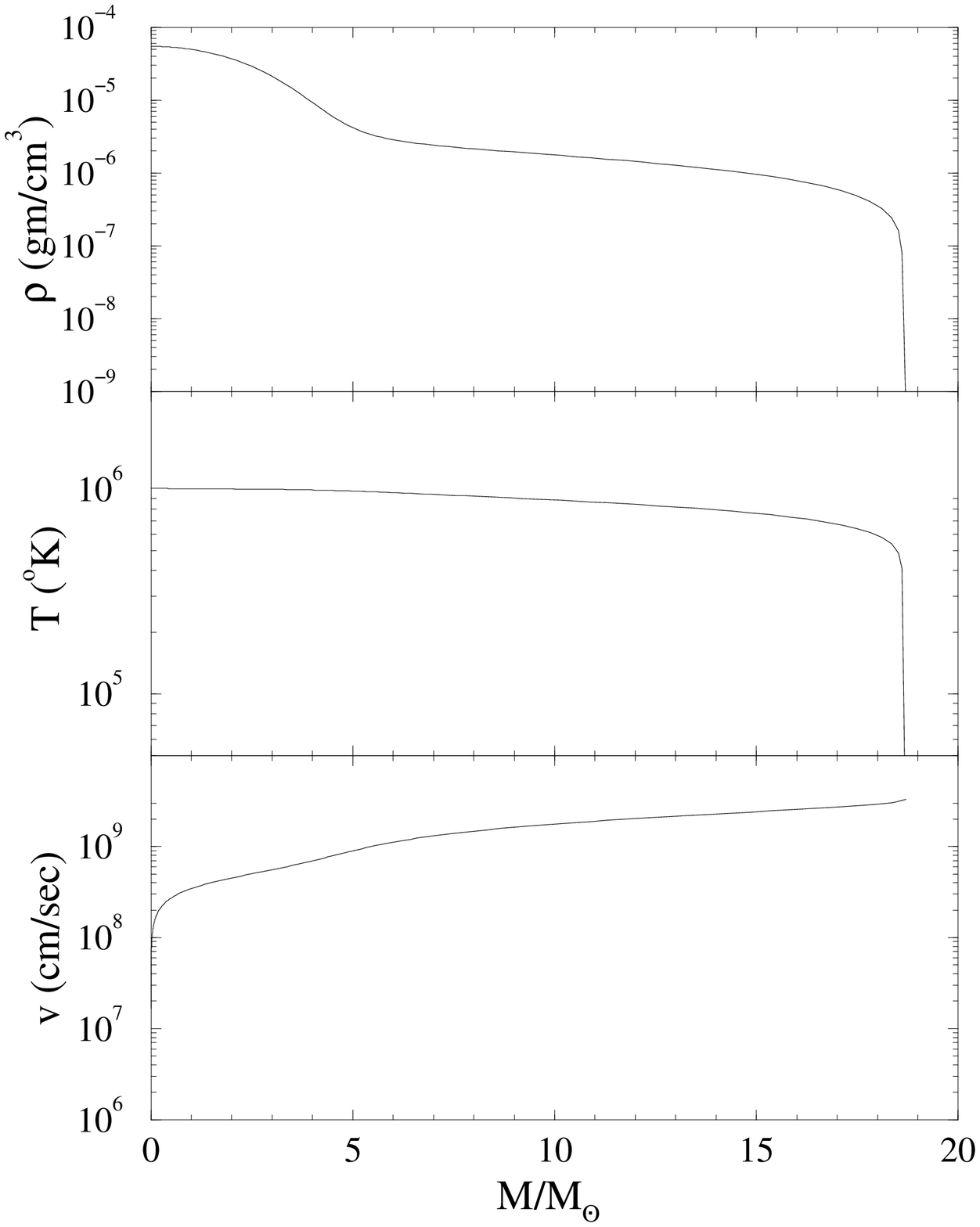}
\figcaption{Initial profile for the toy model:  
density, temperature and velocity.} \label{fig:toymodel}
\end{center}
\end{minipage}
\hspace{2.5cm}
\begin{minipage}[t]{70mm}
\begin{center}
\includegraphics[width=8cm]{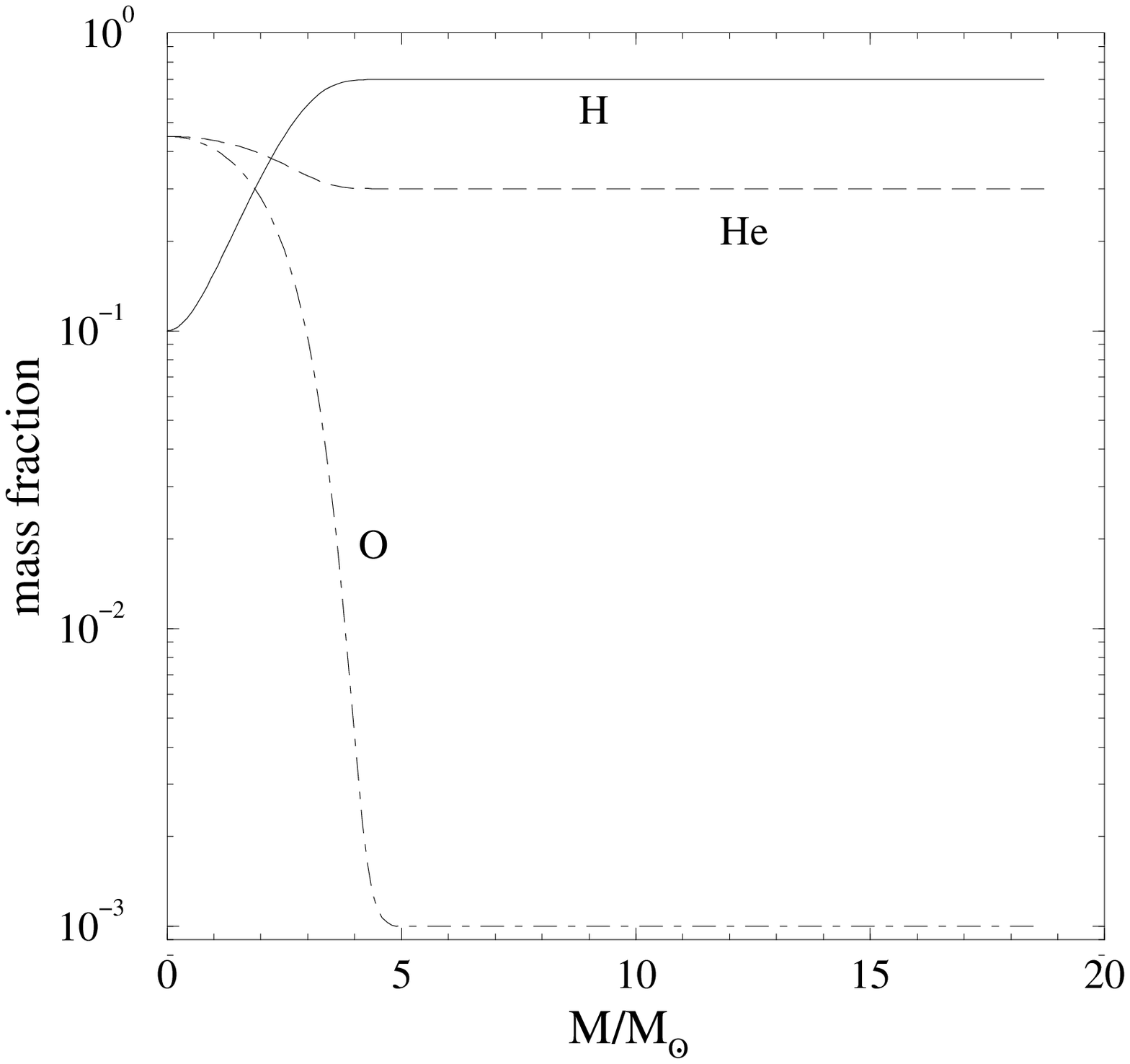}
\figcaption{Initial profile for the toy model: composition (mass fractions 
of H-He-O).}
\label{fig:toycomp}
\end{center}
\end{minipage}
\end{figure*}

We demonstrate here the reliability of our 
rescaling scheme by studying a specific example. We assume a supernova 
envelope somewhat similar to the SN1997D case examined in 
\S~\ref{Sect:SN1997D}, 
with several simplifications that allow for a more rapid computation. The 
basic features of the model are presented in table~\ref{tab:toymodel} and 
in Figs.~\ref{fig:toymodel}-\ref{fig:toycomp}. We note that the initial 
velocity profile throughout the entire envelope was set so that $v(r)=r/t_0$. 
The initial accretion time-scale is determined by the properties of the 
region of bound material. A single fiducial radioactive isotope was included, 
with a mass fraction which is 0.01 of that of oxygen. This isotope 
was assumed to emit all the decay energy in $\gamma-$rays. 

\vspace{15cm}



We have carried out two simulations of this model, the first in its original 
form and once rescaled by a factor $\alpha=5$ (since $t_0\ll t_{acc,0}$ and 
$R_{mb,0}<R_{acc,0}$ to begin with, the model can be rescaled at $t=0$). 
The light curves of both models are compared in 
Fig.~\ref{fig:toylight}, where luminosity and time in the rescaled 
model were rescaled inversely in order to compare them to the results of the 
original model, i.e., $L^s(t^s)\rightarrow \alpha L^s(\alpha^2 t^s),\;$. 
We note that the rescaled model required only 60 CPU hours, compared to 
about 300 CPU hours for the original model to reach an equivalent 
evolutionary stage.

The reliability of rescaling is evident from the very good agreement of the 
light curve and accretion rate history. In particular, the accretion 
luminosities at emergence agree to better than $10\%$ (and a similar 
agreement is found in the accretion rates). We note that there is 
an obvious deviation in the light curve 
during the recombination phase, which is not unexpected since 
recombination is a non-linear process \citep{ArnettBook}: 
the rescaled model has smaller 
physical size, and the recombination front crosses it more quickly. This 
discrepancy is not important for determining the late-time accretion rate 
and luminosity, since the accreting region is unaffected by the recombination 
occurring much further out in the expanding envelope.   

\begin{inlinefigure}
\centerline{\includegraphics[width=0.5\linewidth]{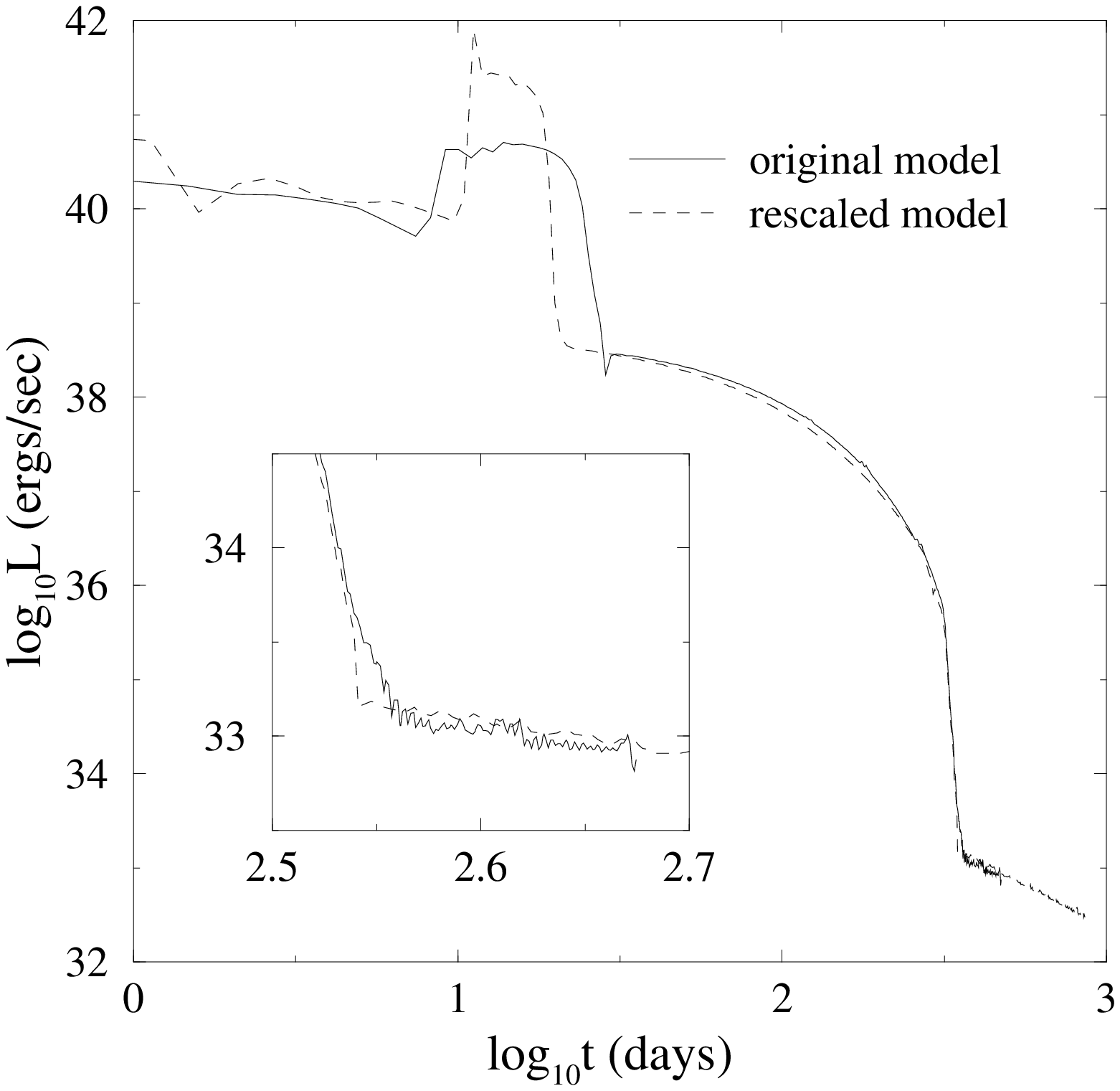}}
\figcaption{Calculated bolometric light curve of the toy model shown in 
Fig.~\ref{fig:toymodel} (solid lines), compared with the calculated light 
curve of model rescaled by a factor of 5 (dashed lines). The inset shows the 
light curves at the time of emergence of the black hole 
\label{fig:toylight}}
\end{inlinefigure}



\begin{thebibliography}{99}

\bibitem[Alexander \& Ferguson(1994)]{AlexFer94}
        Alexander, D. R., \& Ferguson, J. W. 1994,
        \apj, 437, 879

\bibitem[Arnett(1980)]{Arnett80} 
        Arnett, D. 1980, 
        \apj, 237, 541

\bibitem[Arnett(1996)]{ArnettBook}
        Arnett, D. 1996, Supernovae and Nucleosynthesis
        (Princeton: Princeton University Press)

\bibitem[Arnett et al.(1989)]{Arnettal89}
        Arnett, D., Bahcall, J. N., Kirshner, R. P., \& Woosley, S. E. 1989, 
        \araa, 27, 629

\bibitem[Bethe(1990)]{Bethe90}
        Bethe, H. A. 1990, Rev.\ Mod.\ Phys.\, 62, 801

\bibitem[Bethe \& Brown(1995)]{BetheBrown95}
        Bethe, H. A., \& Brown, G. E. 1995, \apjl, 445, L129

\bibitem[Benetti et al.(2000)]{Benetti00}
        Benetti S. et al.\ 2000, \mnras, submitted

\bibitem[Blondin(1986)]{Blondin86} 
        Blondin, J. M. 1986, \apj, 308, 755

\bibitem[Burrows(1988)]{Burrows88}
        Burrows, A. 1988, \apj, 334, 891

\bibitem[Chevalier(1989)]{Chevalier89} 
        Chevalier, R.~A. 1989, 
        \apj, 346, 847

\bibitem[Chevalier(1996)]{Chevalier96} 
        Chevalier, R.~A. 1989, 
        \apj, 459, 322

\bibitem[Chugai \& Utrobin(1999)]{ChugUt99}
        Chugai, N. N., \& Utrobin, V. P. 1999, 
        \aap, 354, 557

\bibitem[Colpi et al.(1996)]{CSW96} 
        Colpi, M., Shapiro, S. L., \& Wasserman, I. 1996, 
        \apj, 470, 1075

\bibitem[Cook et al.(1994)]{CookST94}
        Cook, G. B., Shapiro, S. L., \& Teukolsky, S. A. 1994, 
        \apj, 424, 823

\bibitem[deMello \& Benetti(1997)]{deMello97}
        De Mello, D., \& Benetti, S. 1997, IAUC 6537

\bibitem[Fryer(1999)]{Fryer99}
        Fryer, C. L. 1999, 
        \apj, 522, 413 

\bibitem[Fryer \& Kalogera(1999)]{FryKal99}
        Fryer, C. L., \& Kalogera, V. 1999, 
        \apj, submitted

\bibitem[Fryer et al.(1999)]{Fryeral99}
        Fryer, C. L., Colgate, S. A., \& Pinto, P. A. 1999, 
        \apj, 511, 885

\bibitem[Houck \& Chevalier(1991)]{HouckChev91} 
        Houck, J. C., \& Chevalier, R. A. 1991,\apj, 376, 234

\bibitem[Israelian et al.(1999)]{Israelianal99}
        Israelian, G., et al.\ 1999, \nat, 401, 142

\bibitem[MacFadyen et al.(1999)]{MacFWoosHeg99}
        MacFadyen, A. I., Woosley S. I., \& Heger, A. 
        1999, \apj, submitted  (astro-ph/9910034)        

\bibitem[Magee et al.(1995)]{TOPS}
         Magee, N. H. et al.~ 1995,  
        {\it "Atomic Structure Calculations and New Los Alamos 
         Astrophysical Opacities}, 
         Astronomical Society of the Pacific Conference Series 
         (Astrophysical Applications of Powerful New Databases, 
         S. J. Adelman and W. L. Wiese eds.) 78, 51
         {\tt http://www.t4.lanl.gov/opacity/tops.html} 

\bibitem[Nobili, Turolla \& Zampieri (1991)]{Nobilietal91}
         Nobili, L., Turolla, R., \& Zampieri, L. 1991, \apj, 383, 250

\bibitem[Nomoto et al.(1994)]{Nomotoal94}
        Nomoto, K., Shigeyama, T., Kumagai, S., Yamaoka, H., \& Suzuki, T.
        1994, in Supernovae, Les Houches 1990, eds.~ S.\ A.\ Bludman, 
        R.\ Mochkovitch \& J.\ Zinn-Justin (North Holland), 489.

\bibitem[Patat et al.(1994)]{Patat94}
        Patat, F., Barbon, R., Capparello, E., \& Turatto, M. 1994,
        \aap, 282, 731         

\bibitem[Pinto \& Woosley(1989)]{PinWoos89}
        Pinto, P. A., \&  Woosley, S. E. 1989, \nat, 333, 534

\bibitem[Shigeyama \& Nomoto(1990)]{ShigNom90}
        Shigeyama, T., \& Nomoto, K. 1990,
        \apj, 360, 242

\bibitem[Sollerman et al.(1998)]{Sollerman94W}
        Sollerman, J., Cumming, R. J., \& Lundqvist, P. 1998,
        \apj, 493, 933 

\bibitem[Suntzeff(1998)]{Suntzeff98} 
        Suntzeff, N. B. 1997, in SN1987A: Ten Years After, ed. 
        M. M. Phillips and N. B. Suntzeff (ASP Conference Series).

\bibitem[Timmes et al.(1996)]{TimmesTiCo}
        Timmes, F. X., Woosley, S. A., Hartmann, D. H., 
        \& Hoffman, R. D. 1996, \apj, 464, 332 

\bibitem[Turatto et al.(1998)]{Turatto98}
        Turatto M. et al.~1998, \apj, 498, L129

\bibitem[Woosley(1988)]{Woosley88}
        Woosley, S. E. 1988, \apj, 330, 218

\bibitem[Woosley \& Weaver(1995)]{WoosWeav95}
        Woosley, S. E., \& Waever, T. A. 1995, 
        \apjs, 101, 181

\bibitem[Woosley et al.(1989)]{Woosleyal89}
        Woosley, S. E., Pinto, P. A., \& Hartmann, D. 1989, 
        \apj, 346, 395

\bibitem[Woosley \& Timmes(1996)]{WoosTim96} 
        Woosley, S. E., \& Timmes, F. X. 1996, \nphysa, 606, 137

\bibitem[Young \& Branch(1989)]{YouBra89}
        Young, T. R., \&  Branch, D. 1989, \apj,  342, L79

\bibitem[Young et al.(1998)]{Young98}
        Young T. R., Nomoto, K., Mazalli, P. A., Iwamoto, K., 
        \& Turatto, M. 1998, 
        in Origin of Matter and Evolution of Galaxies '97, 
        ed.~N. Kubono, (World Scientific),  
  
\bibitem[Zampieri et al.(1996)]{Zamp96} 
        Zampieri, L., Miller, J.C., \& Turolla, R. 1996, 
        \mnras, 281, 1183

\bibitem[Zampieri et al.(1998a)]{PaperI}
        Zampieri, L., Colpi, M., Shapiro, S. L., \& Wasserman, I. 1998a, 
        \apj, 505, 876

\bibitem[Zampieri et al.(1998b)]{SN97Dlet}
        Zampieri, L., Shapiro, S. L., \& Colpi, M. 1998b, 
        \apjl, 502, L149


\end{thebibliography}
\end{document}